\documentclass[aps,prb,twocolumn,floatfix,longbibliography]{revtex4-2}
\usepackage{color}
\usepackage{amssymb}
\usepackage{amsmath}
\usepackage{bm}
\usepackage{hyperref}
\usepackage{graphicx}
\usepackage{amsmath,amssymb,bm,epsfig,color}
\usepackage[table]{xcolor}
\usepackage{siunitx}
\usepackage[color=green!60,textsize=small]{todonotes}
\usepackage{enumitem}
\usepackage[normalem]{ulem}

\begin{document}
\title{Polarization Charge around Impurities in Two-Dimensional Anisotropic Dirac Systems}
\author{Mohamed M. Elsayed}
\affiliation{Department of Physics, University of Vermont, Burlington, VT 05405, USA}
\author{Sang Wook Kim}
\affiliation{Department of Physics, University of Vermont, Burlington, VT 05405, USA}
\author{Juan M. Vanegas}
\affiliation{Department of Physics, University of Vermont, Burlington, VT 05405, USA}
\author{Valeri N. Kotov}
\affiliation{Department of Physics, University of Vermont, Burlington, VT 05405, USA}

\begin{abstract}
Introducing quasiparticle anisotropy in graphene via uniaxial strain has a profound effect on the polarization charge density induced by external impurities, both Coulomb and short-range. In particular, the charge distribution induced by a Coulomb impurity exhibits a power law tail modulated by a strain-dependent admixture of angular harmonics. The appearance of distributed charge is in sharp contrast to the response in pristine/isotropic graphene, where for subcritical impurities the polarization charge is fully localized at the impurity position. It is also interesting to note that our results are obtained strictly at zero chemical potential, and the behavior is distinct from the familiar Friedel oscillations observed at finite chemical potential. We find that over a wide range of strain, the $d$-wave symmetry is dominant. The presence  of Dirac cone tilt, relevant to some  2D materials beyond graphene, can also substantially affect the induced charge distribution. Finally we consider impurities with short range potentials, and study the effect of strain on the charge response. Our results were obtained in the continuum  via perturbation theory valid for weak (subcritical) potentials, and supported by numerical lattice simulations based on density functional theory.
\end{abstract}
\maketitle


\section{Introduction}

External impurities in graphene \cite{Antonio2009} and other two-dimensional (2D) electronic  systems provide an important probe into the underlying Dirac quasiparticle 
structure, and allow for the study of effects not present in conventional materials. In the case of external Coulomb impurities, two 
distinct regimes can be identified depending on the impurity strength $Z\alpha$ \cite{Shytov2007,Pereira2007,Kotov2009,Kotov2008,Terekhov2008,Biswas2007}.  For  weak impurities in neutral graphene, in the so-called subcritical regime ($Z\alpha < 1/2$) that is 
accessible perturbatively, the vacuum polarization charge resides on the lattice scale (i.e. in the continuum limit $n({\bf r}) = \tilde{Q} \delta({\bf r})$).
On the other hand, strong impurities in the supercritical regime ($Z\alpha > 1/2$) lead to ``vacuum charging"  characterized by resonances in the quasiparticle density of states  as well
as the appearance of a distributed component in the polarization charge $n({\bf r}) = \tilde{Q} \delta({\bf r}) + A/r^2$. 
These findings have stimulated  extensive experimental studies via scanning tunneling microscopy, yielding results consistent with theoretical predictions\cite{Ovdat2017,Wang2013,Mao2016,Wang2012}.

In this work we investigate how uniaxial strain and Dirac cone tilt affect the polarization charge distribution induced by subcritical impurities in charge neutral graphene.
The effect of uniaxial strain on graphene's electronic structure is well-known \cite{Choi2010,Amorim2016,Pereira2009,Naumis2017,Mohiuddin2009,Roldan2015}.
Under weak strain the Dirac cones become anisotropic, whereas very strong strain (in any direction but armchair) leads to a topological transition towards an insulating state via Dirac cone merger. For simplicity, we focus on uniaxial strain along the armchair direction in our calculations, since the system remains gapless under a wide range of strain values
(ranging from several percent, generally accessible experimentally, to tens of percent, which will be used to illustrate our theoretical results). However, we also briefly consider the case of zigzag strain, and it is evident that the behavior predicted by the armchair model is quite general for strain along any direction, apart from at extreme strains approaching the transition. Another way to break rotational symmetry in momentum space is Dirac cone tilt. This can be realized in deformed graphene, and is naturally present in various 2D materials \cite{Amorim2016, Montambaux2019,Goerbig2008,Zabolotskiy2016,Sadhukhan2017}. Because anisotropy modifies graphene's polarization properties, it generally affects a variety of  phenomena related to interactions with external atoms
such as Kondo impurities, van der Waals interactions with neutral atoms, etc. 
\cite{Sharma2013,DelMaestro2021,Sengupta2018,Kim22}.

For Coulomb impurities we generally find that in the weak coupling regime ($Z\alpha \ll 1$), the presence of strain and/or tilt induced anisotropy leads to the emergence of a distributed polarization charge tail (in addition to the local response), $n({\bf r})  =  \tilde{Q} \delta({\bf r}) +A(\phi)/r^2$. This is in contrast to the case of isotropic graphene, where only a localized charge ($\tilde{Q}$) is present. We analyze in detail the symmetries  and strain dependence of the functions  $\tilde{Q},A(\phi)$, using the conventional continuum random phase approximation (RPA)  for the polarization charge \cite{Simion2005,Giuliani2005}, supplemented by lattice density functional theory (DFT) calculations. We find that the function $A(\phi)$ is oscillatory and the total induced charge is determined by the local component $\tilde{Q}$ (i.e. the net charge carried by the distributed tail is zero, $\int d\phi A(\phi)=0$). We emphasize that our calculations were performed for neutral graphene, at zero chemical potential $\mu =0$. At finite chemical potential $\mu\neq0$ both strained and pristine graphene exhibit Friedel oscillations \cite{Cheianov2006,Wunsch2006,Sadhukhan2017} with a cubic power law tail. It is particularly interesting that in the strict limit $\mu=0$, the radial Friedel oscillations are replaced by the more pronounced quadratic tail with a characteristic angular pattern. Finally, we extend previous work on short-range impurities \cite{Milstein2010,Mkhitaryan2012} to include strain, and find that it produces oscillations in the cubic power law tail.

The rest of the paper is organized as follows. In Section \ref{problem} we summarize the main equations pertaining to graphene's electronic structure and polarization properties in the presence of uniaxial strain. In Section  \ref{coulombimp} we present our results for the external Coulomb impurity problem, including in-plane and off-plane impurities atop uniaxially strained graphene. Section \ref{short} discusses the polarization patterns for short-range impurities. Section \ref{tilt} studies the effect of Dirac cone tilt on the Coulomb impurity response. Section \ref{ZigZag} is a brief study of the Coulomb impurity response under zigzag strain, and
 Section \ref{discussion} contains our conclusions. Appendices \ref{app:substrate}, \ref{app:coulomb}, \ref{app:short}, \ref{app:friedel} summarize some technical details, including  useful functions and integrals, and a brief discussion of Friedel oscillations in anisotropic systems.
 Throughout the paper we set $\hbar=1$.


\section{Problem Formulation: Uniaxial Strain}
\label{problem}

We  study how anisotropy in the Dirac spectrum of graphene
modifies its electrostatic response to charged impurities in, or proximate
to, the plane. We assume graphene at the neutrality point, i.e. the chemical potential $\mu = 0$ throughout
the main text. The spectrum of  Dirac fermions with different velocities ($v_x,v_y$) along the two directions has the form

\begin{equation}
\varepsilon({\bf p})=\pm\sqrt{v_{x}^{2}p_{x}^{2}+v_{y}^{2}p_{y}^{2}},\label{e1}
\end{equation}
and we quantify the anisotropy via ${\displaystyle v\equiv\frac{v_{y}}{v_{x}}}$.
It may be assumed without loss of generality that $v\leq1$. 
We consider a model where strain $\delta$ is applied in the armchair direction ($y$-axis).
In this case the dependence
of  $v_x,v_y$ on strain can be extracted from numerical data \cite{Choi2010} to be

\begin{equation}
v=v(\delta)=\frac{v_{g}(1-2.23\delta)}{v_{g}(1+0.37\delta)}=\frac{v_{y}(\delta)}{v_{x}(\delta)},
\label{velocities}
\end{equation}
where $v_{g}$ is the velocity in unstrained graphene.
It should be noted that experimentally strain of several percent is readily achievable \cite{Naumis2017,Amorim2016} while in principle, it is known that graphene can sustain much higher strains.
Thus we will employ strain as a theoretical parameter, and allow it to take on a wide range of values, in order to illustrate novel phenomena present in the impurity response. For the case of zigzag strain, considered in Section \ref{ZigZag}, the anisotropy becomes strongly non-linear near the topological transition, and will be extracted for specific strain values from the numerical data of Ref. \cite{Choi2010}.

In linear response, the induced charge density is 
\begin{equation}
n({\bf q})=\frac{V_{ext}({\bf q})\Pi(\textbf{q},0)}{1-V_{e\text{-}e}({\bf q})\Pi(\textbf{q},0)},\label{eq:4}
\end{equation}
where we use the static polarization function ${\displaystyle \Pi(\textbf{q},0)}$
for anisotropic Dirac fermions, and ${\displaystyle V_{ext},\,V_{e\text{-}e}}$
are the potential energies associated with the external impurity and
the internal electron-electron interactions respectively. The term
in the denominator accounts for the RPA screening due to electron
interactions in graphene.

The exact expression for the polarization is easily found to be ($N=4$, taking into account the valley and spin degrees of freedom):
\begin{equation}
{\label{e3}}\Pi(\textbf{q},\omega)=-\frac{N}{16v_{x}v_{y}}\frac{v_{x}^{2}q_{x}^{2}+v_{y}^{2}q_{y}^{2}}{\sqrt{v_{x}^{2}q_{x}^{2}+v_{y}^{2}q_{y}^{2}-\omega^{2}}},
\end{equation}
and 
\begin{equation}
V_{e\text{-}e}({\bf q})=V({\bf q})=\frac{2\pi e^{2}}{\kappa q}
\label{eq:ee}
\end{equation}
is the Fourier transform of the Coulomb interaction between
electron pairs, where $\kappa$ is the effective dielectric constant due to the potential presence of a substrate. 
We define a dimensionless coupling constant 
\begin{equation}
\alpha=\alpha(\delta)=\frac{e^{2}}{\kappa v_{x}}=\frac{e^{2}}{\kappa v_{g}(1+0.37\delta)}=\frac{2.2}{\kappa(1+0.37\delta)}
\end{equation}
that controls the strength of Coulomb interactions in strained graphene.


\section{Coulomb Impurity}
\label{coulombimp}

\subsection{In-Plane}
\label{inplane}
From now on we measure the induced charge density $n$ in units of the positive charge $|e|$,
i.e. we calculate and plot the quantity  $n/|e|$, and  for simplicity we set $|e|=1$ in all formulas.

For an external in plane Coulomb impurity with charge ${\displaystyle Z}|e|$, in 
Eq.~(\ref{eq:4}) we have:
\begin{equation}
V_{ext}({\bf q})=Z\frac{2\pi e^{2}}{\kappa q}
\label{eq:ext}
\end{equation}

\begin{figure}[ht]
\begin{centering}
\includegraphics[width=1\columnwidth]{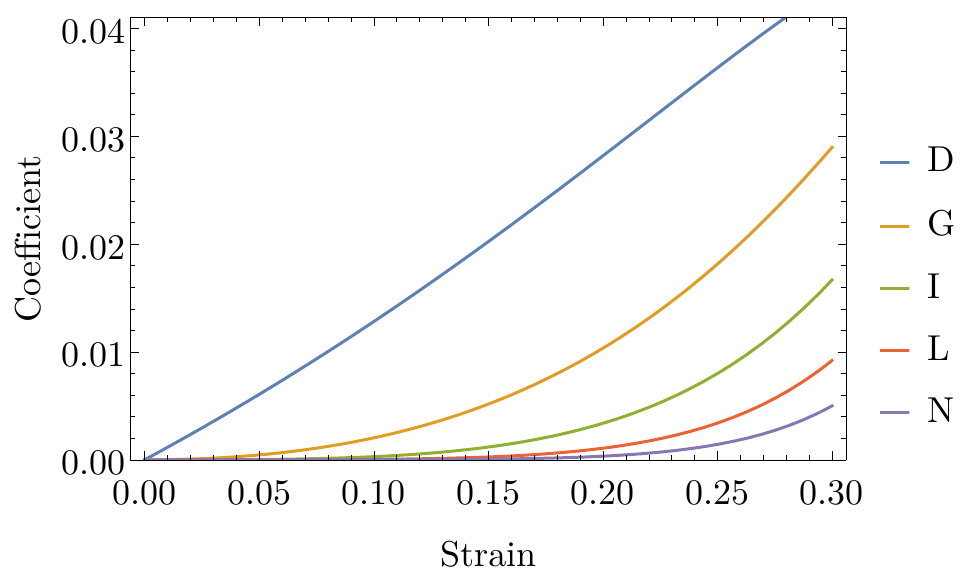} 
\par\end{centering}
\caption{The coefficients $D,G,I,L,N$, appearing in Eq.~(\ref{resultcoulomb}), as a function of strain ${\displaystyle \delta}$
for $\kappa=2.5$.}
\label{v1} 
\end{figure}

In Equations (\ref{eq:ee}) and (\ref{eq:ext}), the effective dielectric constant $\kappa = (1 + \epsilon_s)/2$
takes into account the presence of a substrate with dielectric constant $\epsilon_{s}$ supporting the graphene sheet, and air/vacuum above. Some additional considerations are presented in Appendix \ref{app:substrate}.
For example a typical substrate is SiO$_2$, for which $\kappa \approx 2.5$ ($\epsilon_s \approx 4$); this will be used as the representative value in all of our plots.


\begin{figure}
            
            \begin{minipage}{0.9\linewidth}
                 \includegraphics[width=1\columnwidth]{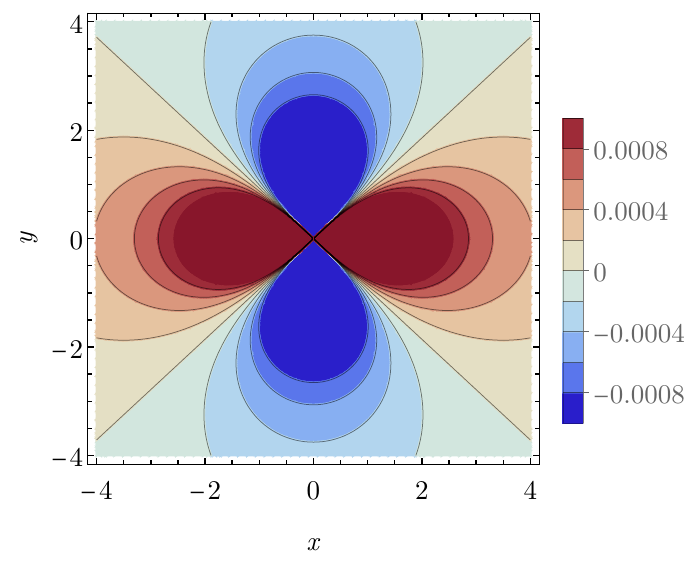}
                
                (a)
                 \vspace{0.4cm}
                \hspace{0.6cm} 
            
        \end{minipage}

        \begin{minipage}{0.9\linewidth}
                 \includegraphics[width=1\columnwidth]{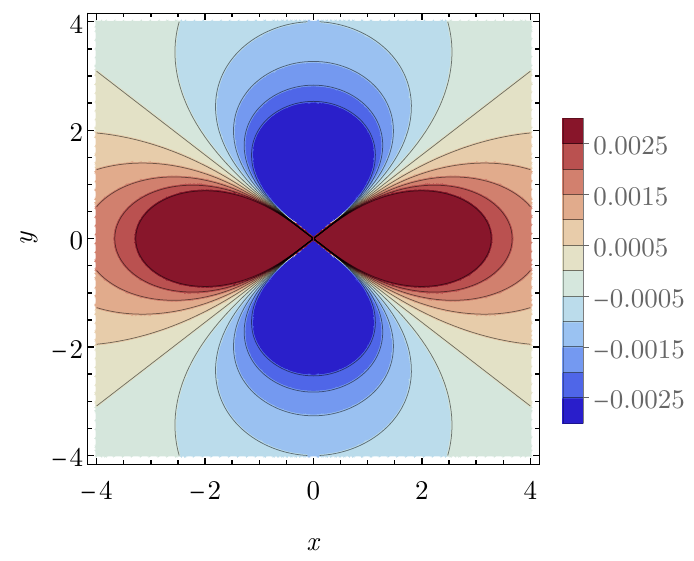}
                  (b)
                \vspace{0.4cm}
                \hspace{0.6cm}
        \end{minipage}  
            
            \begin{minipage}{0.9\linewidth}
                \includegraphics[width=1\columnwidth]{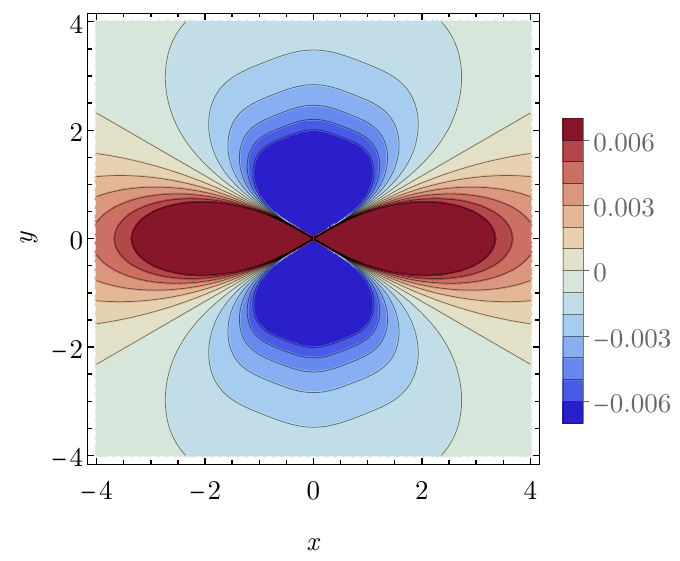}
                (c)
                 \vspace{0.4cm}
                \hspace{0.45cm} 
         \end{minipage}
\caption{Contour plots of the distributed induced charge 
${\displaystyle \tilde{n}_c ({\bf r}) =\frac{n({\bf r})}{Z\alpha}\,+\frac{\pi}{2}\,Q(\delta)\,\delta({\bf r})}$ for
(a) $\delta=0.05$, (b) $\delta=0.15$, and (c) $\delta=0.25$. The coordinates ${\bf r}=(x,y)$ are in arbitrary units as there is no characteristic length scale in the problem. Subsequently, the charge density is measured in the reciprocal of the chosen units squared.
}
\label{CIPplots}
\end{figure}
\begin{figure}
    
    \includegraphics[width=1\columnwidth]{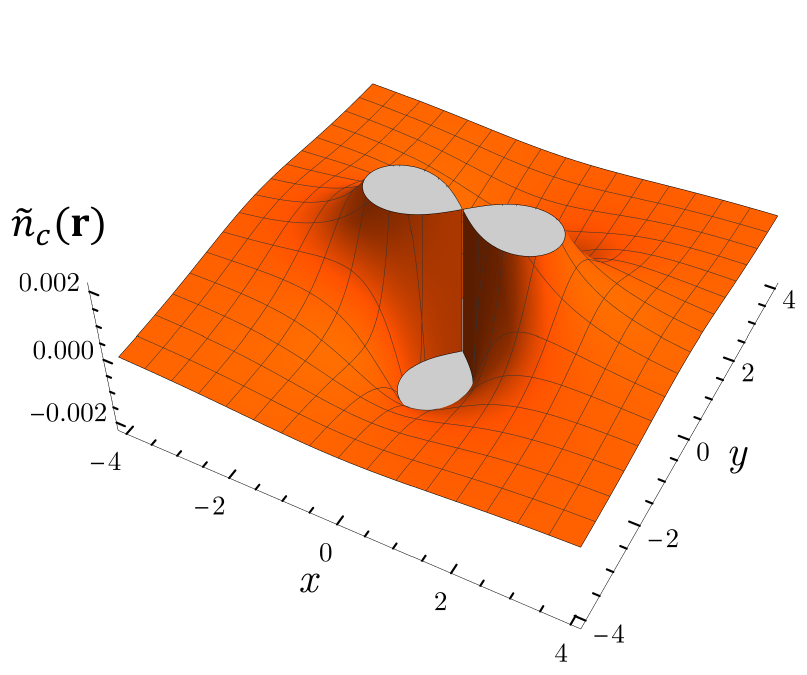}
    \caption{3D density plot of $\tilde{n}_c({\bf r})$ at $\delta=0.05$, in arbitrary units as in Figure \ref{CIPplots}.}
    \label{C5Side}
\end{figure}

\begin{figure*}
            \centering
            \begin{minipage}{0.5\linewidth}
                 \includegraphics[width=1\columnwidth]{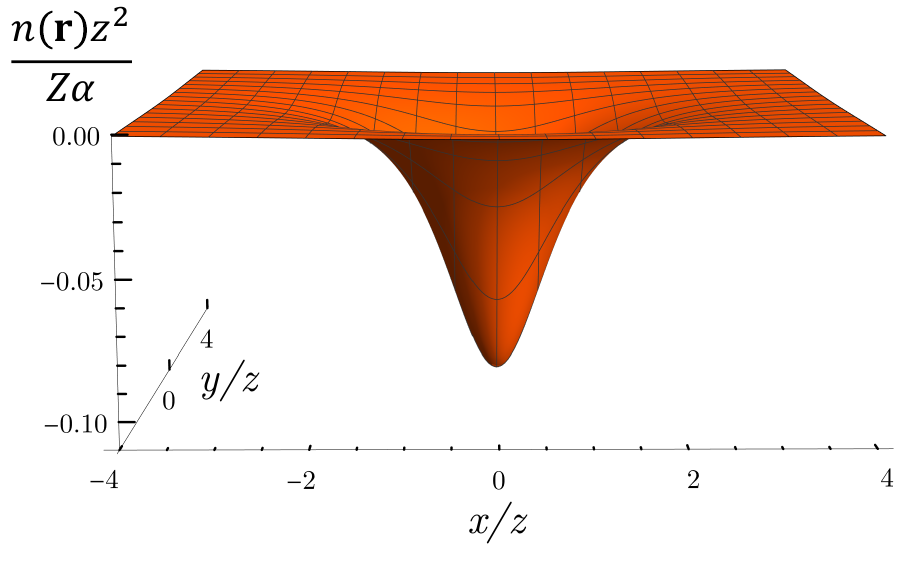}
                 \vspace{0cm}
                \hspace{0.8cm} (a)
        \end{minipage}   
        \hspace{-0.2cm}
        \begin{minipage}{0.5\linewidth}
                 \vspace{-1cm}
                 \includegraphics[width=1\columnwidth]{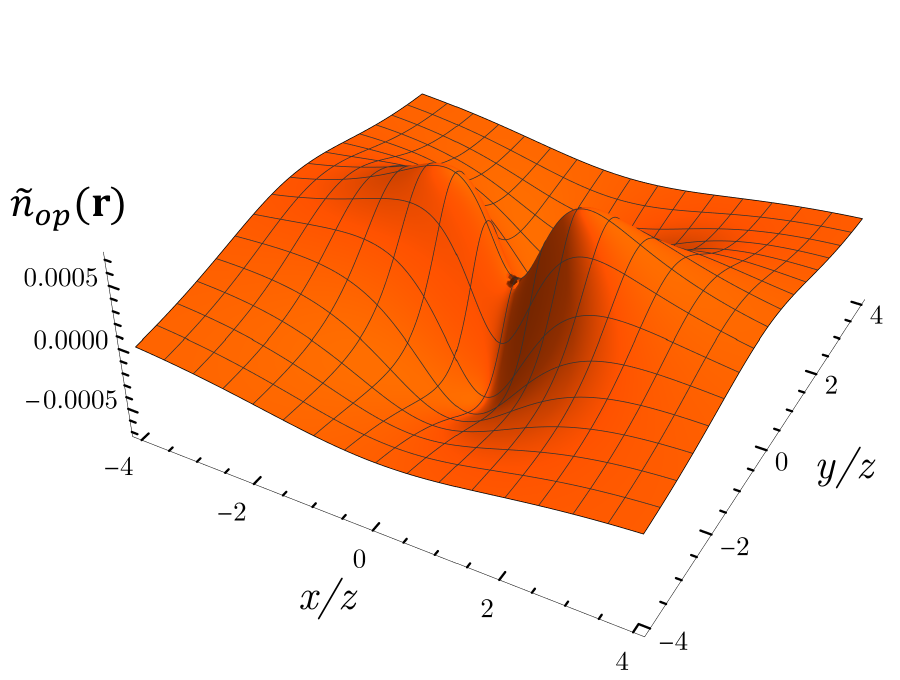}
                 \vspace{0.2cm}
                \hspace{0.7cm} (b)
        \end{minipage}

\caption{Density plots in coordinates $(x/z,y/z)$.
(a) $\frac{n({\bf r})z^{2}}{Z\alpha}$ at $\delta=0.05$. On this scale the density is completely
dominated by the azimuthally symmetric $s$-wave component which is present
in both unstrained and strained graphene. (b) Plot of $\tilde{n}_{op}({\bf r})=\frac{n({\bf r})z^{2}}{Z\alpha}+\frac{Q(\delta)}{4}\frac{1}{[1+(r/z)^{2}]^{3/2}}$,
which is the angularly dependent density induced by strain. The $d$-wave symmetry is dominant, as in the case of the in-plane impurity.}
\label{v45} 
\end{figure*}

Transforming Eq.~(\ref{eq:4}) to real space we find the induced charge
density: 
\begin{equation}
n({\bf r})=-Z\alpha\frac{1}{8\pi v}\int d^{2}{\bf k}\ e^{i{\bf k}\cdot{\bf r}}\ \frac{\sqrt{\cos^{2}\theta+v^{2}\sin^{2}\theta}}{1+\frac{\pi}{2}\frac{\alpha}{v}\sqrt{\cos^{2}\theta+v^{2}\sin^{2}\theta}},\label{eg}
\end{equation}
where ${\displaystyle \theta}$ is the polar angle of ${\displaystyle {\bf k}}$.

It is easy to see from Eq.~(\ref{eg}) that one can extract the results
by use of the identity 
\begin{equation}
e^{i{\bf k}.{\bf r}}=J_{0}(kr)+2\sum_{m=1}^{\infty}i^{m}J_{m}(kr)\cos m(\theta-\phi),
\end{equation}
where $\phi$ is the polar angle of ${\bf r}$.
By parity, we find that the only
terms that contribute to the integral are

\begin{equation}
J_{0}(kr)+2\sum_{m=even}i^{m}J_{m}(kr)\cos(m\theta)\cos(m\phi),
\end{equation}
which is to be substituted back into Eq.~(\ref{eg}). 

It is clear that the first term alone produces the delta function response
while all others contribute to the distributed tail.

Evaluation of the integral yields 
\begin{eqnarray}
 \frac{n({\bf r})}{Z\alpha} & = & -\frac{\pi}{2}Q(\delta)\delta({\bf r})+\frac{1}{r^{2}}\left\{ D(\delta)\cos2\phi \right.\nonumber \\
 &  & \left.+\:G(\delta)\cos4\phi+ I(\delta)\cos6\phi \right.\nonumber \\ 
 & & \left.+\:L(\delta)\cos8\phi + N(\delta)\cos10\phi + \ldots\right\}, \label{et}
 \label{resultcoulomb}
\end{eqnarray}
where the coefficients $Q,D,G,...$ depend on strain and are calculated in  Appendix \ref{app:coulomb}.
This is a perturbative first order result (weak potential) that is formally valid in the limit $Z\alpha \ll 1$.

An important feature of Eq.~(\ref{resultcoulomb}) is that strain has led to the appearance of an oscillatory distributed charge tail $\sim 1/r^2$, in addition to the purely local response found in pristine graphene. Naturally, the coefficients $D,G,\ldots$ vanish at zero strain and we may recover the isotropic result. Notice that the polarization charge (integrated charge density) carried
by the tail is zero and thus the total polarization charge is determined by the local 
component $ \int d^2{\bf r} \frac{n({\bf r})}{Z\alpha} = -\frac{\pi}{2}Q(\delta)$.
The partial wave expansion converges quite nicely, as evidenced by the behavior of the coefficients shown in Figure \ref{v1}. We have omitted $Q(\delta)$ for clarity since it is an order of magnitude greater than the other terms, but the general dependence is that it steadily grows with strain. Figure \ref{CIPplots}
shows contour plots of the distributed tail of the induced charge for different
values of strain. Clearly the $d$-wave angular component is dominant, but note how increased anisotropy more prominently admixes the higher harmonics. For the benefit of the reader, we also include a 3D density plot at an achievable strain of $\delta=0.05$ in Figure \ref{C5Side}.
The strength of the response sensitively depends upon the values of $\delta$ and $\kappa$. Generally speaking, higher strains induce more variation and larger amplitudes in the charge density. However, the influence of $\kappa$ is more subtle: with increasing $\kappa$ the screening from the graphene electrons is reduced ($\alpha$ is smaller), which considerably increases the values of all coefficients $Q,D,G,\ldots\,$; on the other hand, the dimensionless coupling $Z\alpha$ also decreases, which reduces the magnitude of the density variations and also improves the validity of the perturbative approach.

\begin{figure*}[ht]
\begin{centering}
\includegraphics[width=2\columnwidth]{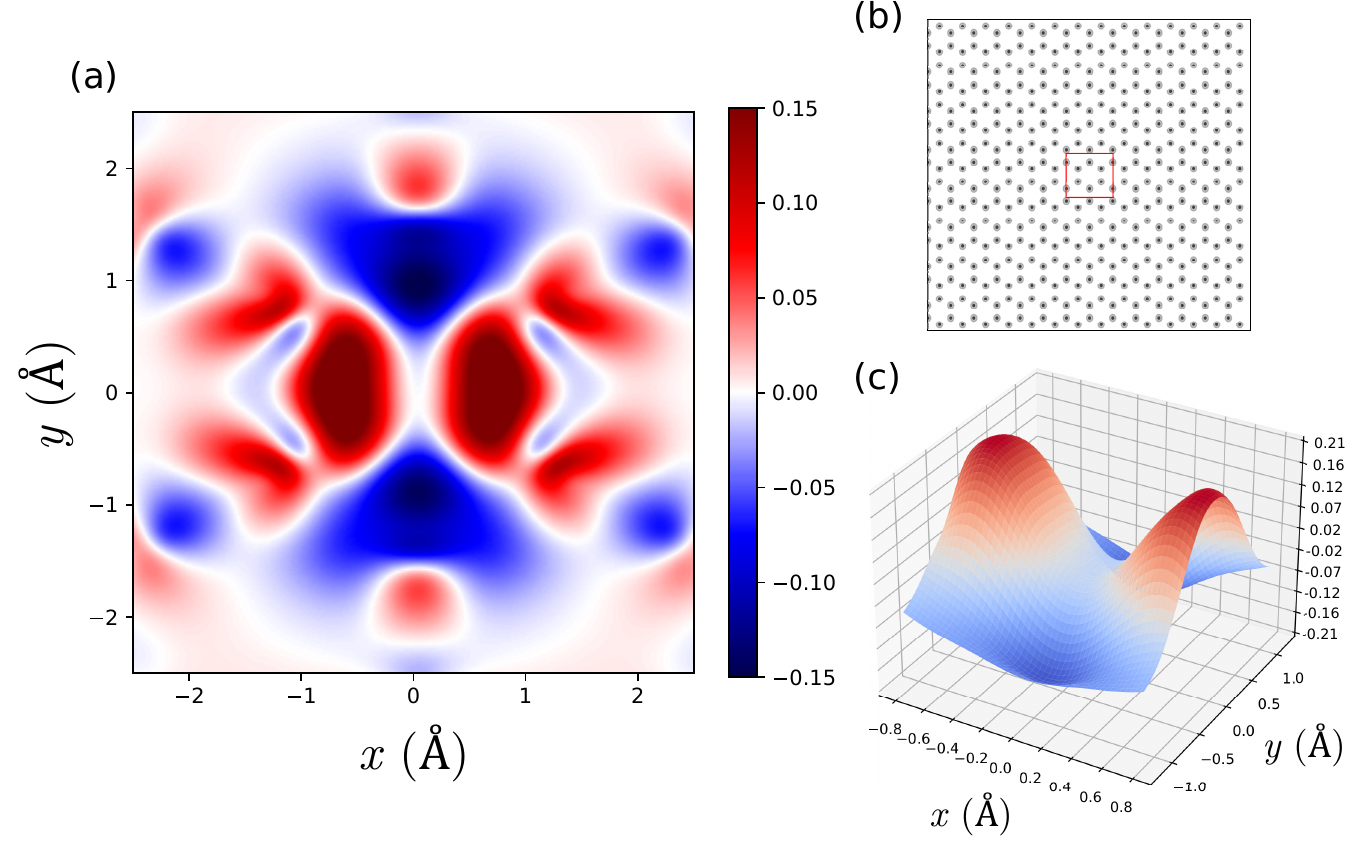}
\par\end{centering}
\caption{(a) Two-dimensional map of the polarization charge density around a Coulomb impurity, obtained by DFT. Panel (b) shows the size of the graphene sheet in our simulations and the area window (in red)  plotted in panel (a). Distances are measured in \AA.
Panel (c) is the corresponding  3D view of the  polarization charge density
near the impurity position  for strain ${\displaystyle \delta=0.2}$. To exclude the isotropic peak and extract the oscillatory tail, we mapped the strained data to the original size of the graphene sheet before subtracting the normalized unstrained data. Thus the figure shows only
the distributed, angularly dependent component of the charge density.}
\label{dft} 
\end{figure*}

\begin{figure}
\begin{centering}
\includegraphics[width=1\columnwidth]{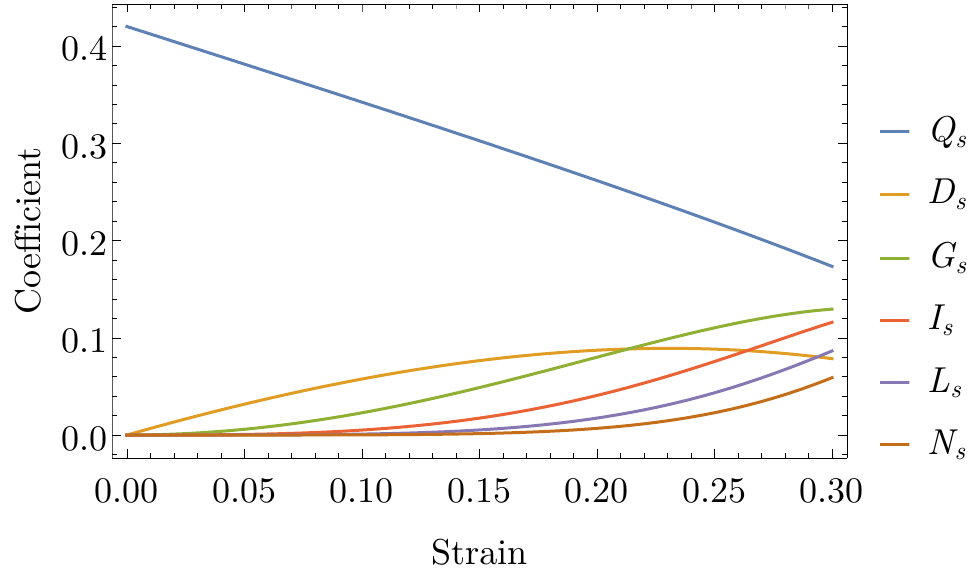} 
\par\end{centering}
\caption{The coefficients $Q_{s},D_{s},G_{s},I_{s},J_{s},K_{s}$, appearing in Eq.~(\ref{coeffshort}), as a function
of strain $\delta$ for $\kappa=2.5$. Explicit formulas are presented in Appendix \ref{app:short}.}
\label{v4} 
\end{figure}

\begin{figure*}
            \centering
            \begin{minipage}{0.33\textwidth}
                 \includegraphics[width=1\textwidth]{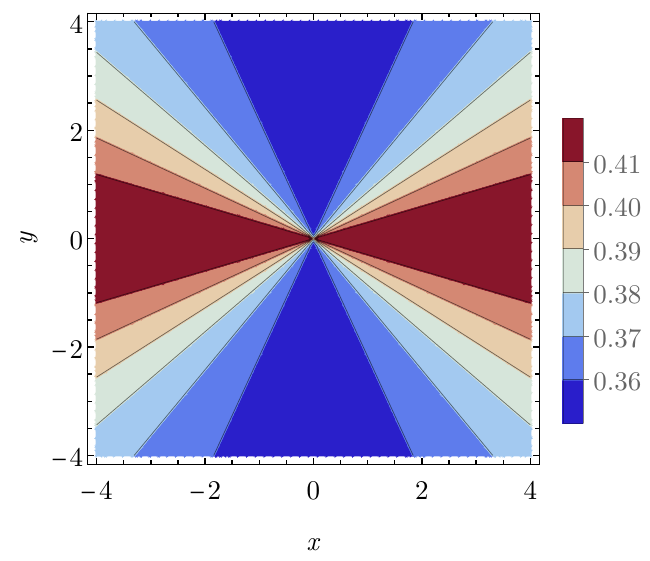}
                 \vspace{0.2cm}
                \hspace{-0.3cm} (a)
        \end{minipage}   
        \hspace{-0.2cm}
        \begin{minipage}{0.33\textwidth}
                 
                 \includegraphics[width=1\textwidth]{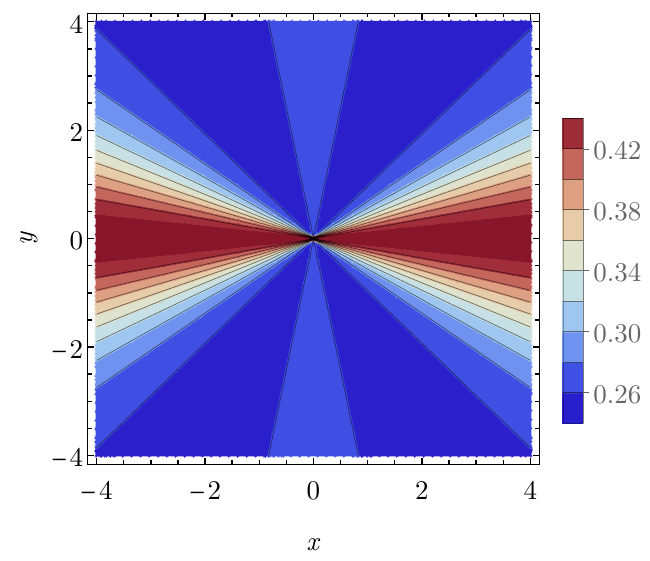}
                 \vspace{0.2cm}
                \hspace{-0.3cm} (b)
        \end{minipage}
        \begin{minipage}{0.33\textwidth}
                 \includegraphics[width=1\textwidth]{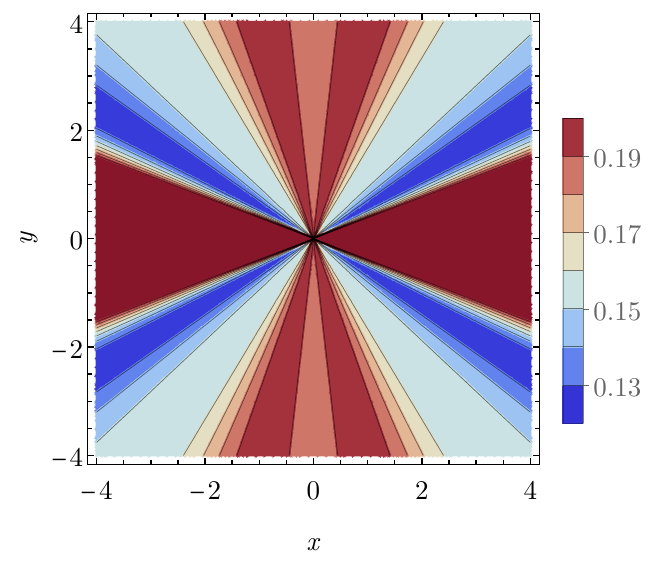}
                 \vspace{0.2cm}
                \hspace{-0.43cm} (c)
        \end{minipage}   
\caption{Contour plots of the function $F(\phi)\sim n({\bf r})r^{3}$ for (a) $\delta = 0.05$, (b) $\delta=0.15$, (c) $\delta=0.25$.}
\label{v5more} 
\end{figure*}

\begin{figure*}
            \centering
            \begin{minipage}{0.33\textwidth}
                 \includegraphics[width=1\textwidth]{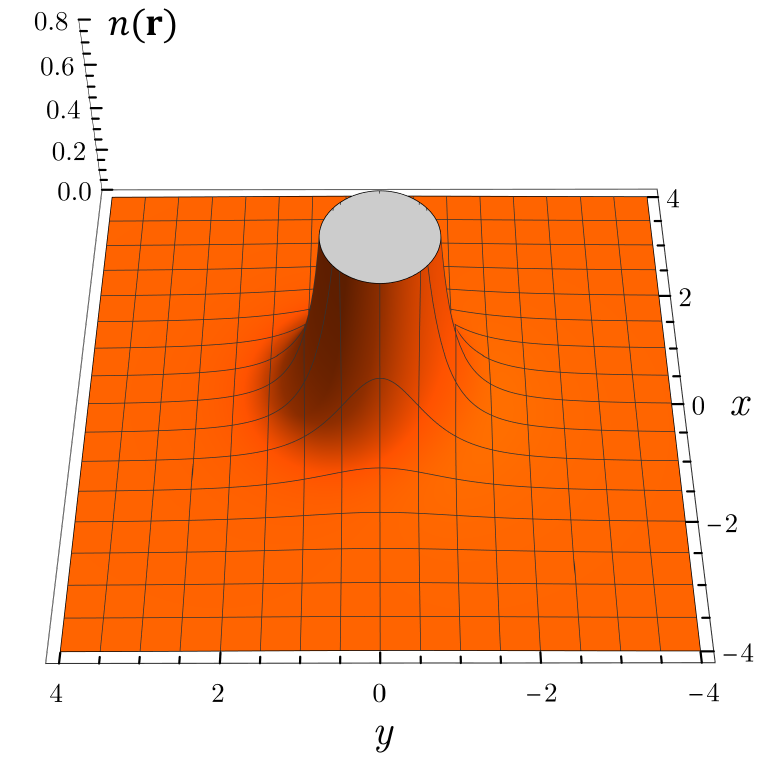}
                 \vspace{0.2cm}
                \hspace{-0.2cm} (a)
        \end{minipage}   
        \hspace{-0.2cm}
        \begin{minipage}{0.33\textwidth}
                 
                 \includegraphics[width=1\textwidth]{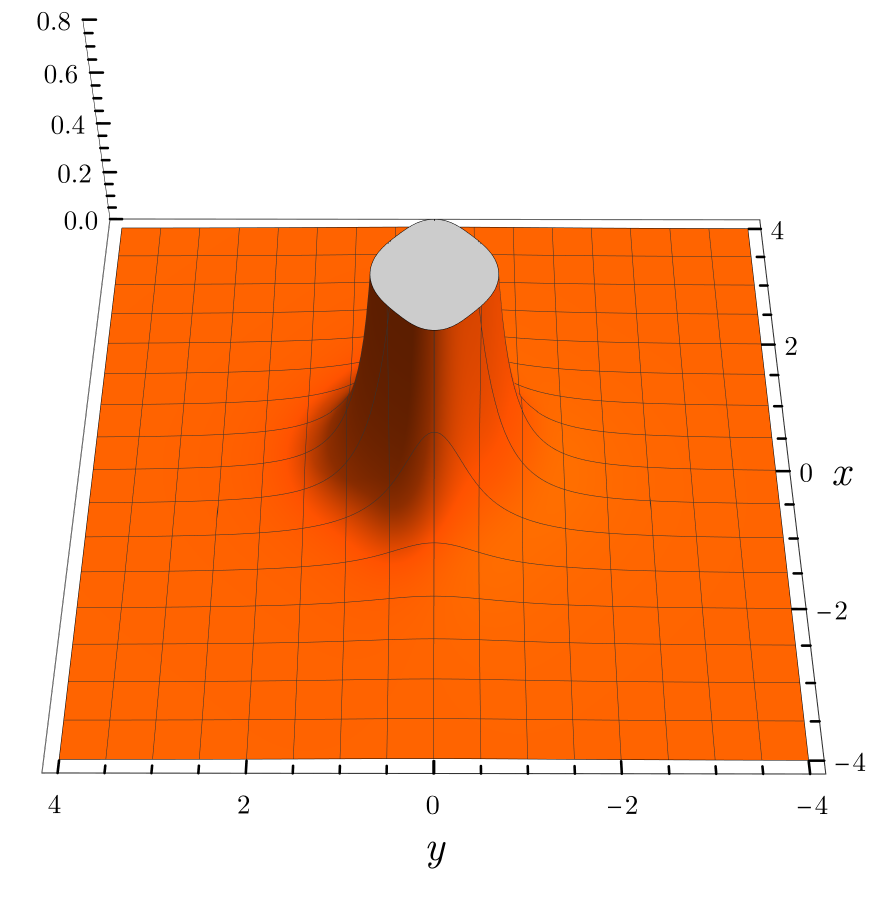}
                 \vspace{0.2cm}
                \hspace{-0.2cm} (b)
        \end{minipage}
        \begin{minipage}{0.33\textwidth}
                 \includegraphics[width=1\textwidth]{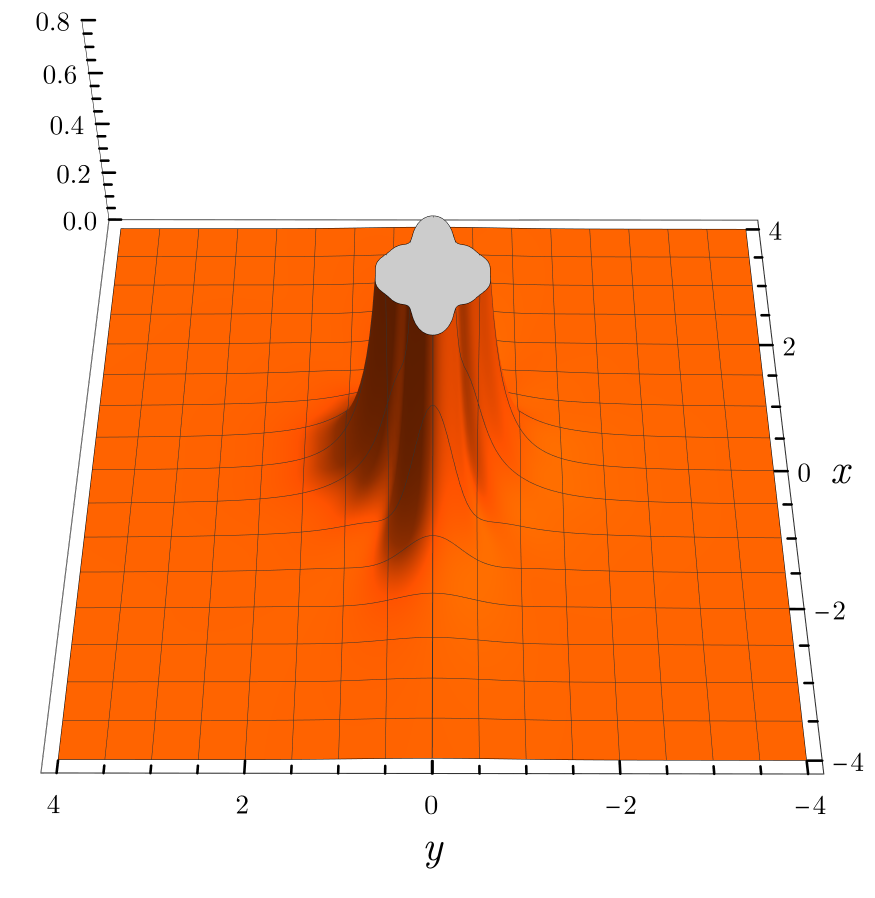}
                 \vspace{0.2cm}
                \hspace{-0.2cm} (c)
        \end{minipage}   
\caption{Evolution of the full density profile $n({\bf r})\sim F(\phi)/r^{3}$
for (a) $\delta = 0.05$, (b) $\delta=0.15$, (c) $\delta=0.25$. Keeping in mind Eq.~(\ref{strainshort})
and restoring the dimensional quantities, the coordinates are measured in units of the characteristic length $U/v_y$,
i.e. $r \rightarrow r/(U/v_y)$,  while the density $n({\bf r})$ is in units of  $v_y^2/(8\pi U^2)$.
}
\label{v5stillmore} 
\end{figure*}


\subsection{Distance $z$ from the Plane}
\label{outplane}

For a Coulomb impurity at a finite distance $z$ from the plane, $V_{ext}(r)=Ze^{2}/(\kappa\sqrt{r^{2}+z^{2}})$,
we have the modified Fourier transform of the external potential:

\begin{equation}
V_{ext}({\bf q})=Z\frac{2\pi e^{2}e^{-qz}}{\kappa q}.
\end{equation}

We perform the calculation again with this simple modification,
measuring all distances in units of the only length-scale $z$. The
final result is:

\begin{multline}
\frac{n({\bf r})z^{2}}{Z\alpha}=-\frac{Q(\delta)}{4}\frac{1}{[1+(r/z)^{2}]^{3/2}}\\
+\frac{1}{(r/z)^{2}}\left\{ \frac{D(\delta)}{2}f_{2}(r/z)\cos2\phi\right.+\frac{G(\delta)}{4}f_{4}(r/z)\cos4\phi\\
+\frac{I(\delta)}{6}f_{6}(r/z)\cos6\phi+\frac{L(\delta)}{8}f_{8}(r/z)\cos8\phi\\
\left.+\frac{N(\delta)}{10}f_{10}(r/z)\cos10\phi\right\}. 
\label{eq:12}
\end{multline}
The functions $f_{2},f_{4},\ldots$ are defined in Appendix \ref{app:coulomb}. Note
that the central ${\displaystyle \delta({\bf r})}$ peak has been
regularized.

By taking the limit of Eq. (\ref{eq:12}) as ${\displaystyle z/r\to0}$
and using
\begin{align*}
\lim_{z/r\to0}\frac{1}{z^{2}[1+(r/z)^{2}]^{3/2}} & =2\pi\delta({\bf r}),\\
\lim_{z/r\to0}f_{n}(r/z) & =n,
\end{align*}
we recover the in plane impurity result Eq.(\ref{et}).

Our results are summarized in Fig. \ref{v45}.
 The main  behavior, including the angular symmetry  of the charge response,
is broadly similar to the case of the in-plane impurity, apart from an overall
smearing and suppression of the sharp features.

\subsection{DFT Simulation}
\label{dftsim}


Density functional theory (DFT) calculations in the local-spin-density (LSD) approximation were performed using the Gaussian and plane waves method (GPW) as implemented in the Quickstep\cite{Quickstep} module of the CP2K software package \cite{CP2Krev}. The PBEsol \cite{PBESol} generalized gradient approximation (GGA), which is based on the PBE (Perdew--Burke--Ernzerhof) \cite{PBE} GGA and optimized for solids and surfaces, was used for the exchange--correlation functional in the DFT calculations. Wavefunction optimization at each self-consistent field (SCF) step was performed with the orbital transformation method \cite{CP2K_OT} and direct inversion in the iterative subspace method. The optimized double-zeta basis set (DZVP-MOLOPT) was used together with the Goedecker–Teter–Hutter (GTH) pseudopotentials \cite{MOLOPT,GTH,GTH2,GTH3}. Graphene was simulated as a periodic sheet consisting of $14 \times 16$ unit cells within a rectangular cell with a vacuum region of $\SI{40}{\angstrom}$. The system is periodic in the plane of the sheet ($x\text{-}y$ plane), but not periodic in $z$. For strained configurations, the system was scaled along the $x$ and $y$ dimensions according to the strain value and the corresponding Poisson ratio (0.165). The Martyna-Tuckerman solver was used for the electrostatic Poisson calculation. 
An external Coulomb potential of a  positive (single) point charge 
was added at a distance  $0.01a$ ($a=\SI{1.42}{\angstrom} $) above  the graphene sheet to compute the effect of the Coulomb impurity. The Quickstep multi-grid was run with a cutoff energy (CUTOFF) of 400 Ry and a relative cutoff (REL\_CUTOFF) of 40 Ry.

The electron density changes due to the Coulomb impurity was obtained by subtracting simulation results without external potential from data including the potential.
Our main goal is to demonstrate the angular effect of uniaxial strain on graphene
and for that purpose we have excluded the first (local) term in Eqs.~(\ref{et}, \ref{eq:12}). We have normalized  the isotropic peak of  the unstrained data and subtracted it from the strained data.


The DFT results are presented in Figure  \ref{dft}.
Notice that the behavior found from the DFT simulations shows a remarkable similarity to  our previous analysis (which was based on a continuum formulation within linear response theory). In particular, the most important feature is the angular variation near the origin
(Figure  \ref{dft}(c)) which  clearly exhibits a dominant $d$-wave like peak and valley behavior  as in Figures \ref{CIPplots}, \ref{v45}.

\begin{figure*}
            \centering
            \hspace{-2.5cm}
            \begin{minipage}{0.5\linewidth}
                 \includegraphics[width=1\columnwidth]{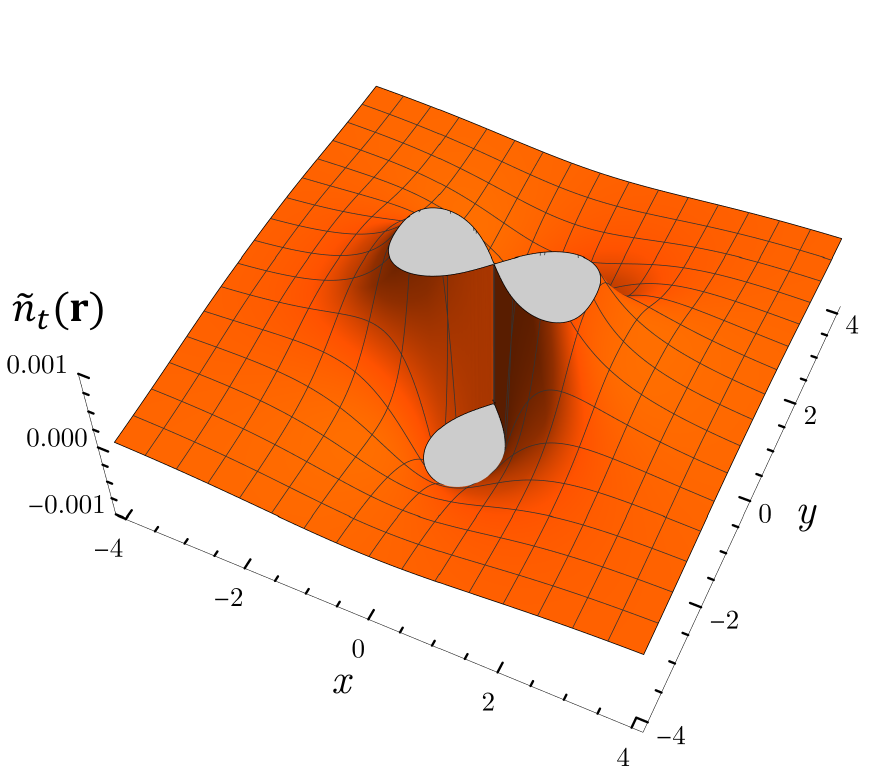}
                 \vspace{0cm}
                \hspace{0cm} (a)
                \hspace{2cm}
        \end{minipage}   
        \hspace{-2cm}
        \begin{minipage}{0.5\linewidth}
                 \vspace{0cm}
                 \includegraphics[width=1\columnwidth]{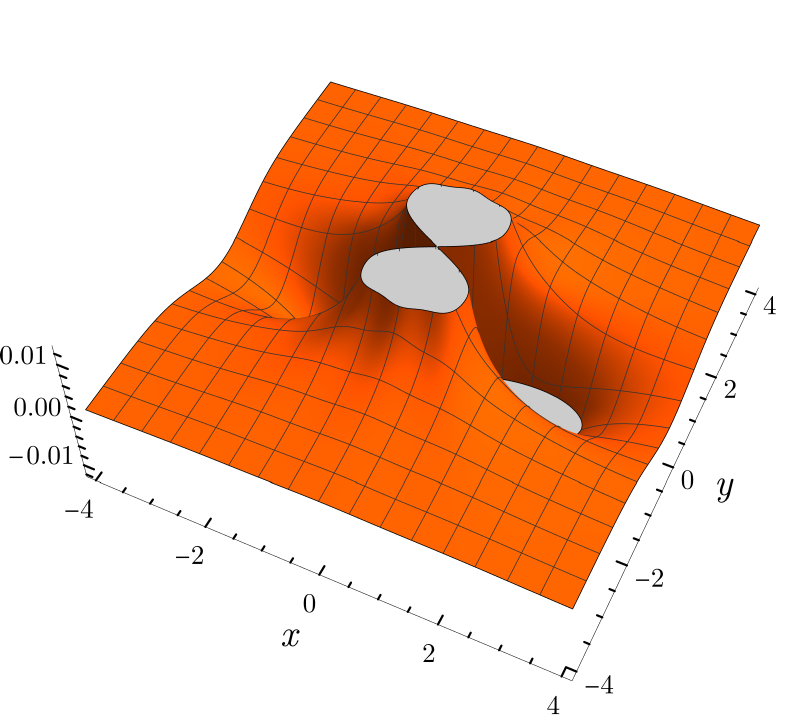}
                 \vspace{0cm}
                \hspace{-6cm} (b)
        \end{minipage}
         \caption{Distributed charge tail defined as in  Fig.~\ref{CIPplots},  for (a) values relevant to borophene, namely
$v=v_{y}/v_{x}=0.8\,(\delta=0.079),\ t=0.46$. The $d$-wave pattern is dominant, similarly to the  response under strain alone. (b) For
$v=v_{y}/v_{x}=0.8,\ t=0.9$, where we have artificially considered
a much larger tilt parameter to observe the effects more clearly. Notice how strong tilt reflects the polarization pattern across the $x\text{-}y$ plane. Coordinates are in arbitrary units, similarly to  Fig.~\ref{CIPplots}.}
\label{fig:tilt}
\end{figure*}

\section{Short-range Impurity}
\label{short}

For an impurity with an interaction range on the scale of graphene's
lattice spacing, one can take:

\begin{equation}
V_{ext}({\bf r})=U\delta({\bf r}).
\end{equation}

This problem has been studied  in detail for  isotropic graphene \cite{Milstein2010,Mkhitaryan2012}. It exhibits a supercritical regime (which requires short distance regularization of the potential),
but we will only consider the weak coupling, subcritical behavior, and more precisely how it is modified by
strain.

Using the  linear response approach, as previously applied to the Coulomb impurity problem,   we have:

\begin{equation}
n({\bf r})=-U\ \frac{1}{4(2\pi)^{2}v_{y}}\int d^{2}ke^{i{\bf k}.{\bf r}}\frac{\vert{\bf k\vert}\sqrt{\cos^{2}\theta+v^{2}\sin^{2}\theta}}{1+\frac{\pi}{2}\frac{\alpha}{v}\sqrt{\cos^{2}\theta+v^{2}\sin^{2}\theta}}.
\end{equation}

This can be represented as

\begin{equation}
n({\bf r})=\frac{1}{r^{3}}\frac{U}{8\pi v_{y}}F(\phi),
\label{strainshort}
\end{equation}
where the function $F(\phi)$ is computed via the partial wave expansion with the result:

\begin{eqnarray}
 F(\phi) & = & Q_{s}(\delta)+D_{s}(\delta)\cos2\phi+G_{s}(\delta)\cos4\phi+I_{s}(\delta)\cos6\phi\nonumber \\
 &  & +\,L_{s}(\delta)\cos8\phi+N_{s}(\delta)\cos10\phi .
 \label{coeffshort}
\end{eqnarray}
The coefficients $Q_{s},D_{s},...$ are defined in  Appendix \ref{app:short}. The
``s'' label denotes short-range.

\begin{figure*}
            \begin{minipage}{0.45\linewidth}
                 \includegraphics[width=1\columnwidth]{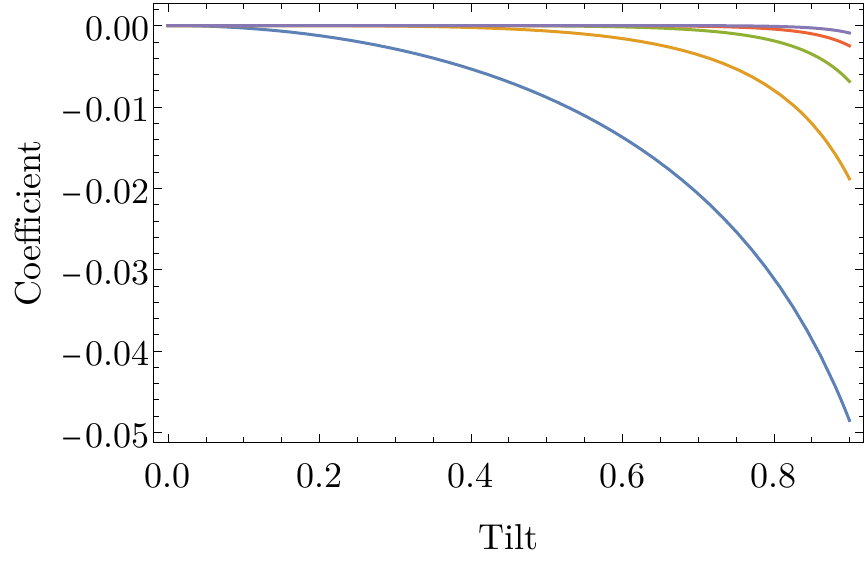}
                 \vspace{-0.2cm}
                \hspace{1.3cm} (a)
        \end{minipage}   
        \hspace{0.2cm}
        \begin{minipage}{0.52\linewidth}
                 \vspace{0cm}
                 \includegraphics[width=1\columnwidth]{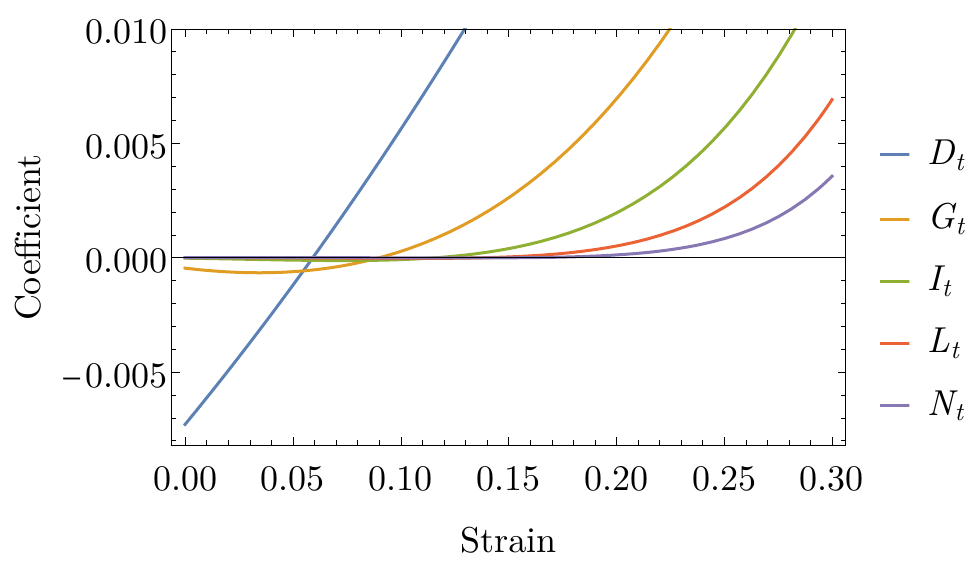}
                 \vspace{0cm}
                \hspace{0.3cm} (b)
        \end{minipage}
         \caption{Plots of the modified coefficients $D_{t}(\delta,t),G_{t}(\delta,t),\ldots$ appearing in Eq.~(\ref{resultcoulomb}). (a) To isolate the effect of tilt, we plot $D_{t}(0,t),G_{t}(0,t),\ldots\;$. (b) Plot of $D_{t}(\delta,0.46),G_{t}(\delta,0.46),\ldots\;$, whereby we fix $t$ at the naturally occurring value in borophene. The competition between the effects of strain $\delta$ and tilt $t$ is clear.}

         \label{TiltCoef}
         \end{figure*}


The primary difference from the in plane Coulomb impurity is that here
the entire induced charge is distributed, and the decay ${\displaystyle \sim1/r^{3}}$
is more rapid. Without strain the function $F(\phi)=constant=Q_{s}$ 
while strain admixes waves with higher symmetries. This
is clear from Fig. \ref{v4} where all the coefficients are plotted as
a function of strain. Note that due to the zero-range nature of the potential, Eq.~(\ref{strainshort}) has a non-integrable singularity at $r=0$ and the total polarization charge diverges. Thus Eq.~(\ref{strainshort}) is only applicable at finite distances beyond a short length scale $a$, which is expected to be on the order of the lattice spacing. The result is relevant for distances $r > a$, and the validity of the perturbative expansion is governed by the condition $\frac{U}{v_ya} \ll 1$, where $U/(v_{y}a)$ is the dimensionless coupling parameter.
Thus overall $r > a \gg U/v_y$.

Figure \ref{v5more} shows the variation of the function $F(\phi)$
for various strains, and Figure \ref{v5stillmore} shows the the density
profile $n({\bf r})$ with the full dependence $\sim1/r^{3}$ included.
For small strain the rapid decay fully obscures the angular variation, but we may easily observe it by plotting $n({\bf r})r^{3}$.
We reach the overall conclusion that the angular dependence for a short range impurity
is more complex than in the Coulomb case, especially for moderate and high strain, and that can be 
traced to the behavior shown in Figure \ref{v4}. The oscillatory terms are several times larger and more evenly weighted, and peculiarly, the isotropic $Q_{s}(\delta)$ component is a decreasing function of strain. On the other hand the angular oscillation is
concomitant with the rather fast $1/r^3$ decay which makes it somewhat less detectable. Note that the counterintuitive  antiscreening  sign
of the polarization charge density  (Eq.~(\ref{strainshort})) is an artifact of the delta function potential.
Upon regularization, one finds that the expected screening charge appears
at distances smaller than the characteristic short-distance regularization scale,
and the total charge contained in the distribution
is zero. This model may be realized
as a point impurity substituted into a graphene lattice site instead
of a carbon atom.



\section{Effect of Dirac cone tilt on the Coulomb Impurity response}
\label{tilt}

We also study the effect of Dirac cone tilt in addition to the strain-induced anisotropy in the Coulomb case. This can manifest in a variety of materials beyond strained graphene \cite{Sadhukhan2017,Zabolotskiy2016,Goerbig2008,Montambaux2019}.
In this case the Hamiltonian has the structure 
\begin{equation}
\hat{H}=v_{x}k_{x}\hat{\sigma}_{x}+v_{y}k_{y}\hat{\sigma}_{y}+tv_{y}k_{y}\hat{\sigma}_{0},
\end{equation}
where the dimensionless parameter $t$ quantifies the degree of cone tilt.
The polarization is calculated to be: 
\begin{equation}
\Pi(\textbf{q},\omega=0)=-\frac{N}{16v_{x}v_{y}}\frac{v_{x}^{2}q_{x}^{2}+v_{y}^{2}q_{y}^{2}}{\sqrt{v_{x}^{2}q_{x}^{2}+v_{y}^{2}(1-t^{2})q_{y}^{2}}}.
\end{equation}
Note that the range of tilt  must be restricted to $t<1$ since a change of band structure occurs at $t=1$ via a Lifshitz transition, whereby the Fermi surface topology changes and nodal lines appear in the spectrum. 
The previously derived formulas in Section \ref{inplane} remain the same with the substitution

\begin{equation}
\varepsilon(\theta)\rightarrow\tilde{\varepsilon}(\theta)=\frac{\sqrt{\cos^{2}\theta\,+v^{2}\sin^{2}\theta}}{1+\frac{\pi}{2}\frac{\alpha}{v}\sqrt{\cos^{2}\theta\,+v^{2}(1-t^{2})\sin^{2}\theta}},
\end{equation}
where the function $\varepsilon(\theta)$ is defined in Eq.~(\ref{epsilondef}), and the corresponding modified coefficients in Eq.~(\ref{resultcoulomb}) become: 
\begin{equation}
    Q(\delta),D(\delta),\ldots\rightarrow Q_{t}(\delta,t),D_{t}(\delta,t),\ldots\;.
\end{equation}

As an example of the relevance of tilt, we look to a member of the borophene family (a 2D boron crystal), for which the naturally occuring parameters are 
\cite{Sadhukhan2017,Zabolotskiy2016}:  $v= v_{y}/v_{x}=0.8,\ t=0.46$. 
We will also consider larger values of tilt to study its effect
more clearly. It has been suggested \cite{Yekta2023} that $t$ may be controlled in borophene by the substitution of carbon atoms into specific lattice sites.

The results are presented in Figure \ref{fig:tilt}. 
We find that there is a competition between the effects of strain and tilt  in the sense that they favor opposite orientations of the dominant 
$d$-wave pattern. For the chosen value of  $\delta \approx 0.1$,  strain produces the primary effect for  $t \lesssim  0.5$,  as shown in 
Figure \ref{fig:tilt}(a),   and tilt has a minor influence.
When tilt plays the primary role  (Figure \ref{fig:tilt}(b) where $t=0.9$) the $d$-wave pattern experiences axis inversion. This can be more precisely understood from the behavior of the coefficients exhibited in Figure \ref{TiltCoef}. It is evident from Figure \ref{TiltCoef}(a), where we plot $D_{t}(0,t),G_{t}(0,t),\ldots\;$, that tilt alone generates a distributed tail, but all the angular oscillations are opposite in sign to those induced by strain. However, as in the response under strain alone, the local component $Q_{t}(0,t)$ is a positive, increasing function of $t$, and an order of magnitude greater than the oscillating terms. The competition between the effects of the two parameters can be distinctly seen in Figure \ref{TiltCoef}(b), where we show the strain dependence at a fixed value of $t=0.46$.

\section{Uniaxial Strain in the Zigzag Direction}
\label{ZigZag}

\begin{figure}
    \begin{minipage}{1\columnwidth}
     \includegraphics[width=0.9\columnwidth]{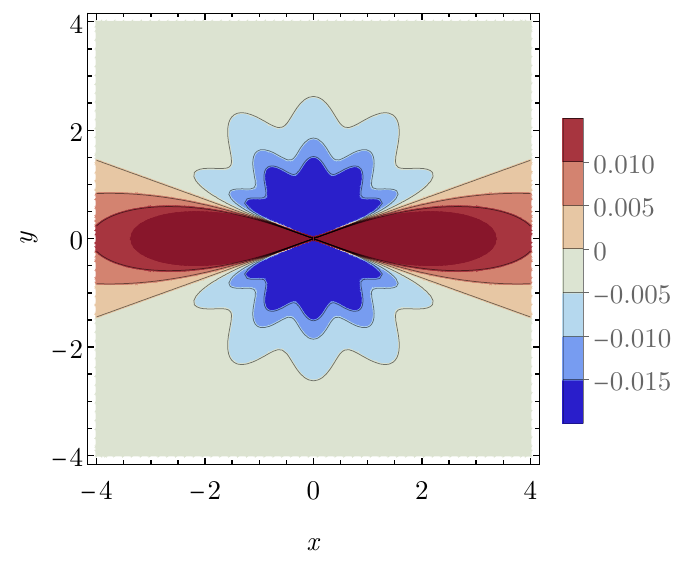}
     \\
     \hspace{-0.7cm}
        (a)
        
    \end{minipage}
    
    \begin{minipage}{1\columnwidth}
        \includegraphics[width=0.9\columnwidth]{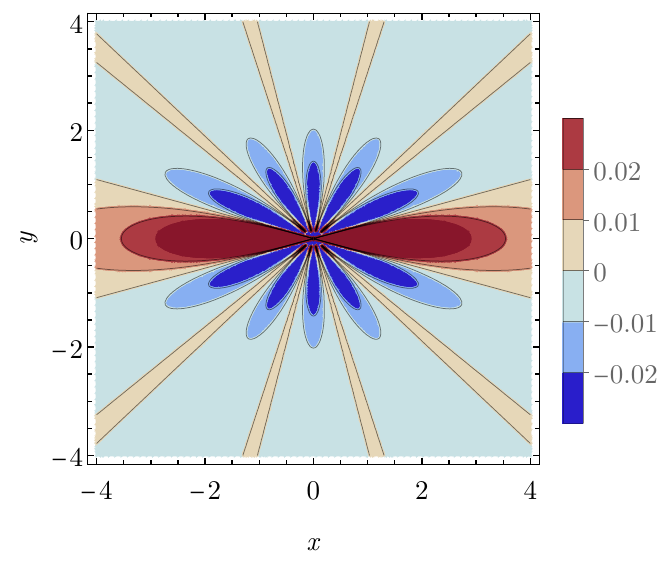}
        \\
        \hspace{-0.55cm}
        (b)
    \end{minipage}
    \caption{Contour plots of the distributed charge density $\tilde{n}_c({\bf r})$ (as defined in Figure \ref{CIPplots}) under  large zigzag ($y$-axis) strains approaching the transition. (a) $\vert \delta-\delta_c\vert\approx0.02$ ($v\approx0.2$), (b) $\vert \delta-\delta_c\vert\approx0.01$ ($v\approx0.08$). }
    \label{ZZContours}
\end{figure}

For completeness, we briefly discuss the in-plane Coulomb impurity response in uniaxially strained graphene along the zigzag direction. The relevant definitions and equations in Sections \ref{problem},\ref{coulombimp}, and Appendix \ref{app:coulomb} remain the same if we rotate the coordinate axes such that the $x$-axis is now the armchair direction and the $y$-axis is the zigzag direction. We adopt this convention throughout this section and in all accompanying figures. 
It is known \cite{Montambaux2019} that under strong enough zigzag strain, graphene approaches a Lifshitz transition characterized by a merger of the Dirac cones and emergence of a semi-Dirac spectrum at the critical point. For small $\delta$, the velocities are slowly varying linear functions, and this can be shown to be true more generally \cite{Naumis2017} for any arbitrary direction of strain. Thus in this regime we expect to see the same behavior as in the armchair case. For strain in the zigzag direction, the linear small $\delta$ behavior extends  to $\delta \approx 0.2$, 
producing  density  patterns virtually identical to the armchair direction. However, as strain increases towards the critical value $\delta_c\approx0.26$ \cite{Choi2010}, $v_y(\delta)$ decreases non-linearly and eventually vanishes, marking the appearance of massive quasiparticles at the transition. (We note that the critical point 
 $\delta_c\approx0.26$ \cite{Choi2010} is slightly larger than the one quoted in \cite{Pereira2009}, $\delta_c\approx0.23$.)

\begin{figure}
  
    \includegraphics[width=1\columnwidth]{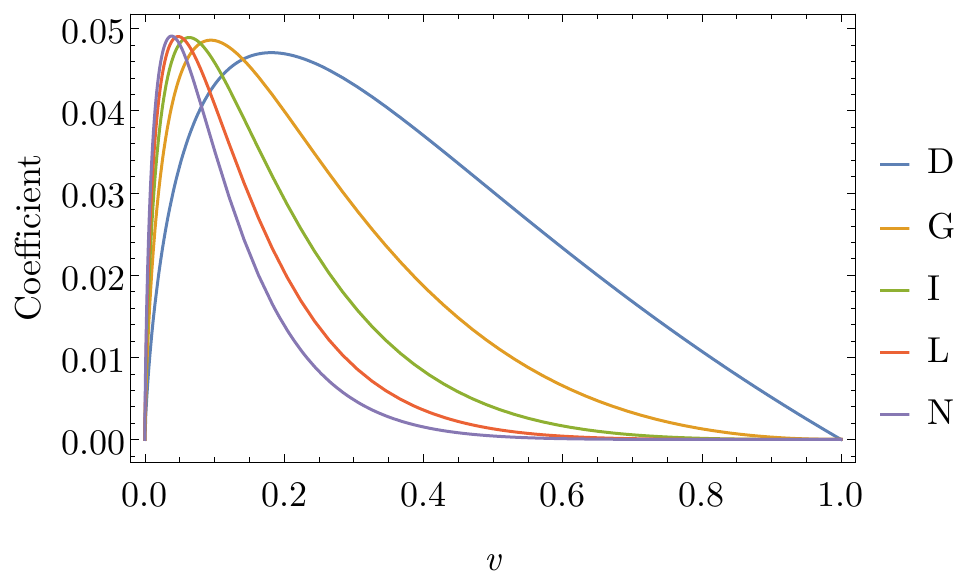}
    \caption{The coefficients in Eq.~(\ref{resultcoulomb}) as a function of the anisotropy $v=v_y/v_x$ for the case of zigzag strain. }
    \label{VHarmonicsPlot}
\end{figure}

In contrast to the milder behavior shown in Eq.~(\ref{velocities}), we now encounter (using the data from Ref.~\cite{Choi2010}) much stronger anisotropy, as quantified by smaller $v$, on the approach to criticality. This causes a dramatic change in the charge response, as can be seen in Figure \ref{ZZContours}. At $\vert\delta-\delta_c\vert\approx0.02$ ($v\approx0.2$), the distribution is still $d$-wave-like but is heavily modulated by higher harmonics; and at $\vert\delta-\delta_c\vert\approx0.01$ ($v\approx0.08$) the signature $d$-wave character is gone altogether and higher symmetry terms become dominant. This can be more clearly understood from the plot of the coefficients as a function of $v$ shown in Figure \ref{VHarmonicsPlot}
(in this plot we use an average value of $\alpha(\delta)$, since it only depends on the slowly-varying velocity in the direction perpendicular to strain). As $v\rightarrow 0$ (maximum anisotropy) every term reaches a maximum, in order of increasing frequency, and starts to decay in such a way that each harmonic has a window where it is the leading component before it is overtaken by the next term. The convergence of the partial wave expansion no longer holds in this regime, and in principle we must retain all terms in Eq.~(\ref{resultcoulomb}) as $v$ tends to zero. Note that this is not problematic, since our model is not valid for $v$ arbitrarily close to zero. There is a crossover at which the behavior of the system is now described by weakly interacting quasi-1D chains, and the 2D Dirac Hamiltonian no longer captures the correct physics. We emphasize that the leading $d$-wave behavior persists up to values of $\delta$ very close to the transition, and this is consistent with our claim about the generality of the results in Section \ref{coulombimp}.

\section{Summary and Discussion}
\label{discussion}
In this paper we have studied the effects of uniaxial strain and Dirac
cone tilt on the vacuum polarization charge induced by subcritical
impurities in charge neutral graphene. In the case of the Coulomb
impurity, the anisotropy produces an unconventional distributed tail
${\displaystyle \sim1/r^{2}}$ in addition to the standard lattice-scale response found in pristine graphene. It has been established elsewhere
that a variety of long range tails can also be generated in the subcritical regime by including  additional interaction
effects or introducing a gap, and are plainly
present in the supercritical regime  \cite{Kotov2009}. 
In the current work  we find that the tail exhibits angular oscillations and carries zero net screening charge, in contrast to the cases mentioned above. Our results show that the $d$-wave symmetry is clearly dominant in a wide range of strain values. 
Stronger anisotropy -- achieved via increasing strain and/or tilt
-- promotes the admixture of higher harmonics, and produces a more
pronounced pattern overall. Thus experimental measurement of the polarization
charge tail on the angstrom scale and beyond, accessible via scanning tunneling
microscopy \cite{Ovdat2017,Wang2013,Mao2016,Wang2012}, may serve as a probe of the degree of mechanical strain
or electronic cone tilt present in a sample.  For example, experiments have successfully detected 
the presence of supercritical  impurity effects. We also note that the
subcritical (perturbative) regime should in principle be more amenable to laboratory conditions given
that the coupling $\alpha$ is suppressed by the presence
of dielectric substrates and electron-electron interactions. In addition, based on the results of Section \ref{outplane}, 
the spatial extent of the angular polarization pattern can  potentially be controlled by the impurity distance from the graphene plane.

It is important to note that our results were obtained at the charge neutrality
point (zero chemical potential). The oscillatory behavior induced by strain and/or tilt is drastically different from  the usual Friedel oscillations at finite $\mu$, briefly discussed in 
 Appendix \ref{app:friedel}. The radial Friedel oscillations occur on the length-scale 
 $1/k_F \sim v_x/|\mu|$ and decay as a cubic power law \cite{Cheianov2006,Wunsch2006,Sadhukhan2017}, in contrast to the angular oscillations we find at strictly zero chemical potential where $n\sim 1/r^2$ with no characteristic length scale.

In the case of the short range impurity potential the effect is less
profound, and the isotropic ${\displaystyle \sim1/r^{3}}$ response
is simply modified by angular oscillations. We again find that increasing
strain favors the mixing of higher symmetries. The zero-range nature of the external potential leads to artifacts in the polarization charge density, namely a non-integrable singularity at the origin, and a counterintuitive antiscreening sign. These issues are resolved by short-distance regularization of the delta function potential.

In our calculations we have mainly focused on a specific model based on armchair uniaxial strain,
but we emphasize that the overall behavior is quite general and would appear for strain in any
direction, with the difference manifesting in the details of the charge polarization patterns. This is supported by the results of Section \ref{ZigZag}, where we study the Coulomb impurity response under zigzag strain, and find that we must consider strains extremely close to the topological transition to find a departure from the signature $d$-wave character. In this regime we do indeed find richer angular variation in the charge density, but it only appears in a small window very close to the transition where the accuracy of the 2D model starts to fail and the system is better described by weakly interacting quasi-1D chains.

\acknowledgments
The authors gratefully acknowledge the financial support from NASA Grant No. 80NSSC19M0143.

\appendix


\section{Impurity Response with Substrate}
\label{app:substrate}

It is known that correctly taking into account the  electrostatics of the substrate-graphene-air/vacuum structure gives the following form for
the induced charge around a Coulomb impurity \cite{Wunsch2006,Radovic2008,Pyatkovskiy2009}
\begin{equation}
n({\bf r})= Z|e|\int \frac{d^{2}{\bf k}}{(2\pi)^2}\ e^{i{\bf k}\cdot{\bf r}} \left\{ \frac{1}{\mathcal{E}({\bf k})}-1 \right \},
\end{equation}
where 
$\mathcal{E}({\bf k}) = \kappa  - V_0({\bf k})\Pi({\bf k})$. Here $V_0({\bf k}) = 2\pi e^2/k$ is the bare Coulomb potential  and $\kappa = (1 + \epsilon_s)/2$ is the effective dielectric constant. $V({\bf k}) = V_0({\bf k})/\kappa$ is the dielectrically screened potential.
The terms cam be rearranged in the following  way:
\begin{eqnarray}
\frac{1}{\mathcal{E}({\bf k})}-1 &=& \frac{V({\bf k})\Pi({\bf k})}{1-V({\bf k})\Pi({\bf k})} + \frac{(1-\kappa)/\kappa}{1-V({\bf k})\Pi({\bf k})}  \nonumber \\
&=& \frac{V({\bf k})\Pi({\bf k})}{\kappa \{1-V({\bf k})\Pi({\bf k})\}} + (1-\kappa)/\kappa,
\label{eq:subrpa}
\end{eqnarray}
which implies that the widely used ``effective" way of treating the substrate, implicit in the RPA  Equations (\ref{eq:4}), (\ref{eq:ee}), (\ref{eq:ext}), somewhat underestimates the localized charge contribution due to the term  $(1-\kappa)/\kappa = (1-\epsilon_s)/(1 + \epsilon_s)$, while it overestimates the distributed tail by a factor of $\kappa$. For example for $\kappa = 2.5$ we have $(1-\kappa)/\kappa =-0.6$ leading to small shifts in the induced density scale (unimportant from the point of view of the present work.)

\section{Coulomb Impurity}
\label{app:coulomb}
The following integrals and  functions appear in the main text of the Section discussing 
in plane impurity.
 
\begin{equation}
\int_{0}^{\infty}dkkJ_{n}(k)=n,\ \ n=2,4,6,8,10
\end{equation}

\begin{equation}
\label{epsilondef}
\varepsilon(\theta,\delta)\equiv\frac{\sqrt{\cos^{2}{\theta}+v^{2}\sin^{2}{\theta}}}{1+\frac{\pi}{2}\frac{\alpha}{v}\sqrt{\cos^{2}{\theta}+v^{2}\sin^{2}{\theta}}}
\end{equation}

\begin{equation}
v=v(\delta)=\frac{1-2.23\delta}{1+0.37\delta}
\end{equation}
\begin{equation}
\alpha=\alpha(\delta)=\frac{2.2}{\kappa(1+0.37\delta)}
\end{equation}

\begin{eqnarray}
 &  & Q(\delta)=\frac{1}{v2\pi}\int_{0}^{2\pi}d\theta\ \varepsilon(\theta,\delta)\\
 &  & D(\delta)=\frac{1}{v2\pi}\int_{0}^{2\pi}d\theta\ \cos(2\theta)\ \varepsilon(\theta,\delta)\\
 &  & G(\delta)=\frac{-1}{v\pi}\int_{0}^{2\pi}d\theta\ \cos(4\theta)\ \varepsilon(\theta,\delta)\\
 &  & I(\delta)=\frac{3}{v2\pi}\int_{0}^{2\pi}d\theta\ \cos(6\theta)\ \varepsilon(\theta,\delta)\\
 &  & J(\delta)=\frac{-2}{v\pi}\int_{0}^{2\pi}d\theta\ \cos(8\theta)\ \varepsilon(\theta,\delta)\\
 &  & K(\delta)=\frac{5}{v2\pi}\int_{0}^{2\pi}d\theta\ \cos(10\theta)\ \varepsilon(\theta,\delta)\\
\nonumber
\end{eqnarray}

For a Coulomb impurity off the plane the following functions appear:

\begin{equation}
f_{n}(r/z)=\int_{0}^{\infty}ds\ s\,e^{-s/(r/z)}\,J_{n}(s)
\end{equation}

 


\section{Short-Range Impurity}
\label{app:short}

The following integrals and  functions appear in the main text of the Section discussing  
the case of short range  impurity.

\begin{equation}
\int_{0}^{\infty}dk\,k^{2}J_{n}(k)=n^{2}-1
\end{equation}

\begin{eqnarray}
Q_{s}(\delta) & = &\frac{1}{2\pi}\int_{0}^{2\pi}d\theta\ \varepsilon(\theta,\delta)\\
 D_{s}(\delta) & = &\frac{6}{2\pi}\int_{0}^{2\pi}d\theta\ \cos(2\theta)\ \varepsilon(\theta,\delta)\\
G_{s}(\delta) & = &\frac{-30}{2\pi}\int_{0}^{2\pi}d\theta\ \cos(4\theta)\ \varepsilon(\theta,\delta)\\
 I_{s}(\delta) & = &\frac{70}{2\pi}\int_{0}^{2\pi}d\theta\ \cos(6\theta)\ \varepsilon(\theta,\delta)\\
J_{s}(\delta) & = &\frac{-126}{2\pi}\int_{0}^{2\pi}d\theta\ \cos(8\theta)\ \varepsilon(\theta,\delta)\\
 K_{s}(\delta)& = &\frac{198}{2\pi}\int_{0}^{2\pi}d\theta\ \cos(10\theta)\ \varepsilon(\theta,\delta)\\
 \nonumber
\end{eqnarray}

\begin{figure}
 \centering
\includegraphics[width= 1\columnwidth]{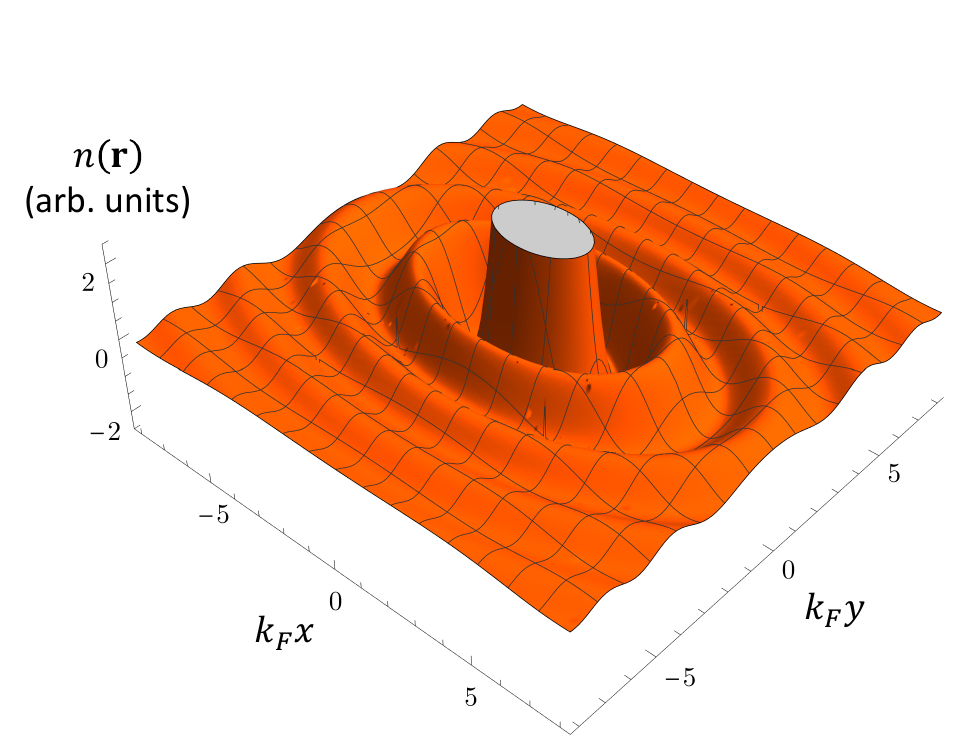}
  \caption{Typical induced charge pattern (in arbitrary units) showing Friedel oscillations for uniaxially strained graphene ($\delta = 0.2$), at finite chemical potential.}
  \label{friedel}
\end{figure}

\section{Anisotropic Friedel Oscillations}
\label{app:friedel}

Consider  finite chemical potential $\mu$.
Then, the result for the polarization is  \cite{Kotov2009}
\begin{equation}
\Pi({\bf q}) = -\frac{2|\mu|}{\pi v_x v_y} + \frac{|\varepsilon({\bf q})|}{2\pi v_x v_y}
G(2|\mu|/|\varepsilon({\bf q})|)
\Theta(|\varepsilon({\bf q})| - 2|\mu|),
\end{equation}
where
\begin{equation}
G(x) = x\sqrt{1-x^2} - \arccos{(x)}.
\end{equation}

We have evaluated numerically the corresponding charge density profile around a Coulomb impurity and the results are presented
in Figure \ref{friedel}, where we define $k_F = |\mu|/v_x$. This is the anisotropic generalization of the
usual Friedel oscillations present in isotropic graphene \cite{Wunsch2006}. The anisotropy of Friedel oscillations in the presence of tilt and strain was pointed out in \cite{Sadhukhan2017}.


\nocite{apsrev42Control}
\bibliographystyle{apsrev4-2}
\bibliography{refs}

\begin{thebibliography}{43}%
\makeatletter
\providecommand \@ifxundefined [1]{%
 \@ifx{#1\undefined}
}%
\providecommand \@ifnum [1]{%
 \ifnum #1\expandafter \@firstoftwo
 \else \expandafter \@secondoftwo
 \fi
}%
\providecommand \@ifx [1]{%
 \ifx #1\expandafter \@firstoftwo
 \else \expandafter \@secondoftwo
 \fi
}%
\providecommand \natexlab [1]{#1}%
\providecommand \enquote  [1]{``#1''}%
\providecommand \bibnamefont  [1]{#1}%
\providecommand \bibfnamefont [1]{#1}%
\providecommand \citenamefont [1]{#1}%
\providecommand \href@noop [0]{\@secondoftwo}%
\providecommand \href [0]{\begingroup \@sanitize@url \@href}%
\providecommand \@href[1]{\@@startlink{#1}\@@href}%
\providecommand \@@href[1]{\endgroup#1\@@endlink}%
\providecommand \@sanitize@url [0]{\catcode `\\12\catcode `\$12\catcode
  `\&12\catcode `\#12\catcode `\^12\catcode `\_12\catcode `\%12\relax}%
\providecommand \@@startlink[1]{}%
\providecommand \@@endlink[0]{}%
\providecommand \url  [0]{\begingroup\@sanitize@url \@url }%
\providecommand \@url [1]{\endgroup\@href {#1}{\urlprefix }}%
\providecommand \urlprefix  [0]{URL }%
\providecommand \Eprint [0]{\href }%
\providecommand \doibase [0]{https://doi.org/}%
\providecommand \selectlanguage [0]{\@gobble}%
\providecommand \bibinfo  [0]{\@secondoftwo}%
\providecommand \bibfield  [0]{\@secondoftwo}%
\providecommand \translation [1]{[#1]}%
\providecommand \BibitemOpen [0]{}%
\providecommand \bibitemStop [0]{}%
\providecommand \bibitemNoStop [0]{.\EOS\space}%
\providecommand \EOS [0]{\spacefactor3000\relax}%
\providecommand \BibitemShut  [1]{\csname bibitem#1\endcsname}%
\let\auto@bib@innerbib\@empty
\bibitem [{\citenamefont {Castro~Neto}\ \emph {et~al.}(2009)\citenamefont
  {Castro~Neto}, \citenamefont {Guinea}, \citenamefont {Peres}, \citenamefont
  {Novoselov},\ and\ \citenamefont {Geim}}]{Antonio2009}%
  \BibitemOpen
  \bibfield  {author} {\bibinfo {author} {\bibfnamefont {A.~H.}\ \bibnamefont
  {Castro~Neto}}, \bibinfo {author} {\bibfnamefont {F.}~\bibnamefont {Guinea}},
  \bibinfo {author} {\bibfnamefont {N.~M.~R.}\ \bibnamefont {Peres}}, \bibinfo
  {author} {\bibfnamefont {K.~S.}\ \bibnamefont {Novoselov}},\ and\ \bibinfo
  {author} {\bibfnamefont {A.~K.}\ \bibnamefont {Geim}},\ }\bibfield  {title}
  {\bibinfo {title} {The electronic properties of graphene},\ }\href
  {https://doi.org/10.1103/RevModPhys.81.109} {\bibfield  {journal} {\bibinfo
  {journal} {Rev. Mod. Phys.}\ }\textbf {\bibinfo {volume} {81}},\ \bibinfo
  {pages} {109} (\bibinfo {year} {2009})}\BibitemShut {NoStop}%
\bibitem [{\citenamefont {Shytov}\ \emph {et~al.}(2007)\citenamefont {Shytov},
  \citenamefont {Katsnelson},\ and\ \citenamefont {Levitov}}]{Shytov2007}%
  \BibitemOpen
  \bibfield  {author} {\bibinfo {author} {\bibfnamefont {A.~V.}\ \bibnamefont
  {Shytov}}, \bibinfo {author} {\bibfnamefont {M.~I.}\ \bibnamefont
  {Katsnelson}},\ and\ \bibinfo {author} {\bibfnamefont {L.~S.}\ \bibnamefont
  {Levitov}},\ }\bibfield  {title} {\bibinfo {title} {Vacuum polarization and
  screening of supercritical impurities in graphene},\ }\href
  {https://doi.org/10.1103/PhysRevLett.99.236801} {\bibfield  {journal}
  {\bibinfo  {journal} {Phys. Rev. Lett.}\ }\textbf {\bibinfo {volume} {99}},\
  \bibinfo {pages} {236801} (\bibinfo {year} {2007})}\BibitemShut {NoStop}%
\bibitem [{\citenamefont {Pereira}\ \emph {et~al.}(2007)\citenamefont
  {Pereira}, \citenamefont {Nilsson},\ and\ \citenamefont
  {Castro~Neto}}]{Pereira2007}%
  \BibitemOpen
  \bibfield  {author} {\bibinfo {author} {\bibfnamefont {V.~M.}\ \bibnamefont
  {Pereira}}, \bibinfo {author} {\bibfnamefont {J.}~\bibnamefont {Nilsson}},\
  and\ \bibinfo {author} {\bibfnamefont {A.~H.}\ \bibnamefont {Castro~Neto}},\
  }\bibfield  {title} {\bibinfo {title} {Coulomb impurity problem in
  graphene},\ }\href {https://doi.org/10.1103/PhysRevLett.99.166802} {\bibfield
   {journal} {\bibinfo  {journal} {Phys. Rev. Lett.}\ }\textbf {\bibinfo
  {volume} {99}},\ \bibinfo {pages} {166802} (\bibinfo {year}
  {2007})}\BibitemShut {NoStop}%
\bibitem [{\citenamefont {Kotov}\ \emph {et~al.}(2012)\citenamefont {Kotov},
  \citenamefont {Uchoa}, \citenamefont {Pereira}, \citenamefont {Guinea},\ and\
  \citenamefont {Castro~Neto}}]{Kotov2009}%
  \BibitemOpen
  \bibfield  {author} {\bibinfo {author} {\bibfnamefont {V.~N.}\ \bibnamefont
  {Kotov}}, \bibinfo {author} {\bibfnamefont {B.}~\bibnamefont {Uchoa}},
  \bibinfo {author} {\bibfnamefont {V.~M.}\ \bibnamefont {Pereira}}, \bibinfo
  {author} {\bibfnamefont {F.}~\bibnamefont {Guinea}},\ and\ \bibinfo {author}
  {\bibfnamefont {A.~H.}\ \bibnamefont {Castro~Neto}},\ }\bibfield  {title}
  {\bibinfo {title} {Electron-electron interactions in graphene: Current status
  and perspectives},\ }\href {https://doi.org/10.1103/RevModPhys.84.1067}
  {\bibfield  {journal} {\bibinfo  {journal} {Rev. Mod. Phys.}\ }\textbf
  {\bibinfo {volume} {84}},\ \bibinfo {pages} {1067} (\bibinfo {year}
  {2012})}\BibitemShut {NoStop}%
\bibitem [{\citenamefont {Kotov}\ \emph {et~al.}(2008)\citenamefont {Kotov},
  \citenamefont {Pereira},\ and\ \citenamefont {Uchoa}}]{Kotov2008}%
  \BibitemOpen
  \bibfield  {author} {\bibinfo {author} {\bibfnamefont {V.~N.}\ \bibnamefont
  {Kotov}}, \bibinfo {author} {\bibfnamefont {V.~M.}\ \bibnamefont {Pereira}},\
  and\ \bibinfo {author} {\bibfnamefont {B.}~\bibnamefont {Uchoa}},\ }\bibfield
   {title} {\bibinfo {title} {Polarization charge distribution in gapped
  graphene: Perturbation theory and exact diagonalization analysis},\ }\href
  {https://doi.org/10.1103/PhysRevB.78.075433} {\bibfield  {journal} {\bibinfo
  {journal} {Phys. Rev. B}\ }\textbf {\bibinfo {volume} {78}},\ \bibinfo
  {pages} {075433} (\bibinfo {year} {2008})}\BibitemShut {NoStop}%
\bibitem [{\citenamefont {Terekhov}\ \emph {et~al.}(2008)\citenamefont
  {Terekhov}, \citenamefont {Milstein}, \citenamefont {Kotov},\ and\
  \citenamefont {Sushkov}}]{Terekhov2008}%
  \BibitemOpen
  \bibfield  {author} {\bibinfo {author} {\bibfnamefont {I.~S.}\ \bibnamefont
  {Terekhov}}, \bibinfo {author} {\bibfnamefont {A.~I.}\ \bibnamefont
  {Milstein}}, \bibinfo {author} {\bibfnamefont {V.~N.}\ \bibnamefont
  {Kotov}},\ and\ \bibinfo {author} {\bibfnamefont {O.~P.}\ \bibnamefont
  {Sushkov}},\ }\bibfield  {title} {\bibinfo {title} {{Screening of Coulomb
  Impurities in Graphene}},\ }\href
  {https://doi.org/10.1103/PhysRevLett.100.076803} {\bibfield  {journal}
  {\bibinfo  {journal} {Phys. Rev. Lett.}\ }\textbf {\bibinfo {volume} {100}},\
  \bibinfo {pages} {076803} (\bibinfo {year} {2008})}\BibitemShut {NoStop}%
\bibitem [{\citenamefont {Biswas}\ \emph {et~al.}(2007)\citenamefont {Biswas},
  \citenamefont {Sachdev},\ and\ \citenamefont {Son}}]{Biswas2007}%
  \BibitemOpen
  \bibfield  {author} {\bibinfo {author} {\bibfnamefont {R.~R.}\ \bibnamefont
  {Biswas}}, \bibinfo {author} {\bibfnamefont {S.}~\bibnamefont {Sachdev}},\
  and\ \bibinfo {author} {\bibfnamefont {D.~T.}\ \bibnamefont {Son}},\
  }\bibfield  {title} {\bibinfo {title} {Coulomb impurity in graphene},\ }\href
  {https://doi.org/10.1103/PhysRevB.76.205122} {\bibfield  {journal} {\bibinfo
  {journal} {Phys. Rev. B}\ }\textbf {\bibinfo {volume} {76}},\ \bibinfo
  {pages} {205122} (\bibinfo {year} {2007})}\BibitemShut {NoStop}%
\bibitem [{\citenamefont {Ovdat}\ \emph {et~al.}(2017)\citenamefont {Ovdat},
  \citenamefont {Mao}, \citenamefont {Jiang}, \citenamefont {Andrei},\ and\
  \citenamefont {Akkermans}}]{Ovdat2017}%
  \BibitemOpen
  \bibfield  {author} {\bibinfo {author} {\bibfnamefont {O.}~\bibnamefont
  {Ovdat}}, \bibinfo {author} {\bibfnamefont {J.}~\bibnamefont {Mao}}, \bibinfo
  {author} {\bibfnamefont {Y.}~\bibnamefont {Jiang}}, \bibinfo {author}
  {\bibfnamefont {E.~Y.}\ \bibnamefont {Andrei}},\ and\ \bibinfo {author}
  {\bibfnamefont {E.}~\bibnamefont {Akkermans}},\ }\bibfield  {title} {\bibinfo
  {title} {Observing a scale anomaly and a universal quantum phase transition
  in graphene},\ }\href {https://doi.org/10.1038/s41467-017-00591-8} {\bibfield
   {journal} {\bibinfo  {journal} {Nature Communications}\ }\textbf {\bibinfo
  {volume} {8}},\ \bibinfo {pages} {507} (\bibinfo {year} {2017})}\BibitemShut
  {NoStop}%
\bibitem [{\citenamefont {Wang}\ \emph {et~al.}(2013)\citenamefont {Wang},
  \citenamefont {Wong}, \citenamefont {Shytov}, \citenamefont {Brar},
  \citenamefont {Choi}, \citenamefont {Wu}, \citenamefont {Tsai}, \citenamefont
  {Regan}, \citenamefont {Zettl}, \citenamefont {Kawakami}, \citenamefont
  {Louie}, \citenamefont {Levitov},\ and\ \citenamefont {Crommie}}]{Wang2013}%
  \BibitemOpen
  \bibfield  {author} {\bibinfo {author} {\bibfnamefont {Y.}~\bibnamefont
  {Wang}}, \bibinfo {author} {\bibfnamefont {D.}~\bibnamefont {Wong}}, \bibinfo
  {author} {\bibfnamefont {A.~V.}\ \bibnamefont {Shytov}}, \bibinfo {author}
  {\bibfnamefont {V.~W.}\ \bibnamefont {Brar}}, \bibinfo {author}
  {\bibfnamefont {S.}~\bibnamefont {Choi}}, \bibinfo {author} {\bibfnamefont
  {Q.}~\bibnamefont {Wu}}, \bibinfo {author} {\bibfnamefont {H.-Z.}\
  \bibnamefont {Tsai}}, \bibinfo {author} {\bibfnamefont {W.}~\bibnamefont
  {Regan}}, \bibinfo {author} {\bibfnamefont {A.}~\bibnamefont {Zettl}},
  \bibinfo {author} {\bibfnamefont {R.~K.}\ \bibnamefont {Kawakami}}, \bibinfo
  {author} {\bibfnamefont {S.~G.}\ \bibnamefont {Louie}}, \bibinfo {author}
  {\bibfnamefont {L.~S.}\ \bibnamefont {Levitov}},\ and\ \bibinfo {author}
  {\bibfnamefont {M.~F.}\ \bibnamefont {Crommie}},\ }\bibfield  {title}
  {\bibinfo {title} {{Observing Atomic Collapse Resonances in Artificial Nuclei
  on Graphene}},\ }\href {https://doi.org/10.1126/science.1234320} {\bibfield
  {journal} {\bibinfo  {journal} {Science}\ }\textbf {\bibinfo {volume}
  {340}},\ \bibinfo {pages} {734} (\bibinfo {year} {2013})}\BibitemShut
  {NoStop}%
\bibitem [{\citenamefont {Mao}\ \emph {et~al.}(2016)\citenamefont {Mao},
  \citenamefont {Jiang}, \citenamefont {Moldovan}, \citenamefont {Li},
  \citenamefont {Watanabe}, \citenamefont {Taniguchi}, \citenamefont {Masir},
  \citenamefont {Peeters},\ and\ \citenamefont {Andrei}}]{Mao2016}%
  \BibitemOpen
  \bibfield  {author} {\bibinfo {author} {\bibfnamefont {J.}~\bibnamefont
  {Mao}}, \bibinfo {author} {\bibfnamefont {Y.}~\bibnamefont {Jiang}}, \bibinfo
  {author} {\bibfnamefont {D.}~\bibnamefont {Moldovan}}, \bibinfo {author}
  {\bibfnamefont {G.}~\bibnamefont {Li}}, \bibinfo {author} {\bibfnamefont
  {K.}~\bibnamefont {Watanabe}}, \bibinfo {author} {\bibfnamefont
  {T.}~\bibnamefont {Taniguchi}}, \bibinfo {author} {\bibfnamefont {M.~R.}\
  \bibnamefont {Masir}}, \bibinfo {author} {\bibfnamefont {F.~M.}\ \bibnamefont
  {Peeters}},\ and\ \bibinfo {author} {\bibfnamefont {E.~Y.}\ \bibnamefont
  {Andrei}},\ }\bibfield  {title} {\bibinfo {title} {Realization of a tunable
  artificial atom at a supercritically charged vacancy in graphene},\ }\href
  {https://doi.org/10.1038/nphys3665} {\bibfield  {journal} {\bibinfo
  {journal} {Nature Physics}\ }\textbf {\bibinfo {volume} {12}},\ \bibinfo
  {pages} {545} (\bibinfo {year} {2016})}\BibitemShut {NoStop}%
\bibitem [{\citenamefont {Wang}\ \emph {et~al.}(2012)\citenamefont {Wang},
  \citenamefont {Brar}, \citenamefont {Shytov}, \citenamefont {Wu},
  \citenamefont {Regan}, \citenamefont {Tsai}, \citenamefont {Zettl},
  \citenamefont {Levitov},\ and\ \citenamefont {Crommie}}]{Wang2012}%
  \BibitemOpen
  \bibfield  {author} {\bibinfo {author} {\bibfnamefont {Y.}~\bibnamefont
  {Wang}}, \bibinfo {author} {\bibfnamefont {V.~W.}\ \bibnamefont {Brar}},
  \bibinfo {author} {\bibfnamefont {A.~V.}\ \bibnamefont {Shytov}}, \bibinfo
  {author} {\bibfnamefont {Q.}~\bibnamefont {Wu}}, \bibinfo {author}
  {\bibfnamefont {W.}~\bibnamefont {Regan}}, \bibinfo {author} {\bibfnamefont
  {H.-Z.}\ \bibnamefont {Tsai}}, \bibinfo {author} {\bibfnamefont
  {A.}~\bibnamefont {Zettl}}, \bibinfo {author} {\bibfnamefont {L.~S.}\
  \bibnamefont {Levitov}},\ and\ \bibinfo {author} {\bibfnamefont {M.~F.}\
  \bibnamefont {Crommie}},\ }\bibfield  {title} {\bibinfo {title} {{Mapping
  Dirac quasiparticles near a single Coulomb impurity on graphene}},\ }\href
  {https://doi.org/10.1038/nphys2379} {\bibfield  {journal} {\bibinfo
  {journal} {Nature Physics}\ }\textbf {\bibinfo {volume} {8}},\ \bibinfo
  {pages} {653} (\bibinfo {year} {2012})}\BibitemShut {NoStop}%
\bibitem [{\citenamefont {Choi}\ \emph {et~al.}(2010)\citenamefont {Choi},
  \citenamefont {Jhi},\ and\ \citenamefont {Son}}]{Choi2010}%
  \BibitemOpen
  \bibfield  {author} {\bibinfo {author} {\bibfnamefont {S.-M.}\ \bibnamefont
  {Choi}}, \bibinfo {author} {\bibfnamefont {S.-H.}\ \bibnamefont {Jhi}},\ and\
  \bibinfo {author} {\bibfnamefont {Y.-W.}\ \bibnamefont {Son}},\ }\bibfield
  {title} {\bibinfo {title} {Effects of strain on electronic properties of
  graphene},\ }\href {https://doi.org/10.1103/PhysRevB.81.081407} {\bibfield
  {journal} {\bibinfo  {journal} {Phys. Rev. B}\ }\textbf {\bibinfo {volume}
  {81}},\ \bibinfo {pages} {081407} (\bibinfo {year} {2010})}\BibitemShut
  {NoStop}%
\bibitem [{\citenamefont {Amorim}\ \emph {et~al.}(2016)\citenamefont {Amorim},
  \citenamefont {Cortijo}, \citenamefont {de~Juan}, \citenamefont {Grushin},
  \citenamefont {Guinea}, \citenamefont {Guti{\'{e}}rrez-Rubio}, \citenamefont
  {Ochoa}, \citenamefont {Parente}, \citenamefont {Rold{\'{a}}n}, \citenamefont
  {San-Jose}, \citenamefont {Schiefele}, \citenamefont {Sturla},\ and\
  \citenamefont {Vozmediano}}]{Amorim2016}%
  \BibitemOpen
  \bibfield  {author} {\bibinfo {author} {\bibfnamefont {B.}~\bibnamefont
  {Amorim}}, \bibinfo {author} {\bibfnamefont {A.}~\bibnamefont {Cortijo}},
  \bibinfo {author} {\bibfnamefont {F.}~\bibnamefont {de~Juan}}, \bibinfo
  {author} {\bibfnamefont {A.~G.}\ \bibnamefont {Grushin}}, \bibinfo {author}
  {\bibfnamefont {F.}~\bibnamefont {Guinea}}, \bibinfo {author} {\bibfnamefont
  {A.}~\bibnamefont {Guti{\'{e}}rrez-Rubio}}, \bibinfo {author} {\bibfnamefont
  {H.}~\bibnamefont {Ochoa}}, \bibinfo {author} {\bibfnamefont
  {V.}~\bibnamefont {Parente}}, \bibinfo {author} {\bibfnamefont
  {R.}~\bibnamefont {Rold{\'{a}}n}}, \bibinfo {author} {\bibfnamefont
  {P.}~\bibnamefont {San-Jose}}, \bibinfo {author} {\bibfnamefont
  {J.}~\bibnamefont {Schiefele}}, \bibinfo {author} {\bibfnamefont
  {M.}~\bibnamefont {Sturla}},\ and\ \bibinfo {author} {\bibfnamefont
  {M.~A.~H.}\ \bibnamefont {Vozmediano}},\ }\bibfield  {title} {\bibinfo
  {title} {{Novel effects of strains in graphene and other two dimensional
  materials}},\ }\href
  {https://doi.org/http://dx.doi.org/10.1016/j.physrep.2015.12.006} {\bibfield
  {journal} {\bibinfo  {journal} {Phys. Rep.}\ }\textbf {\bibinfo {volume}
  {617}},\ \bibinfo {pages} {1} (\bibinfo {year} {2016})}\BibitemShut {NoStop}%
\bibitem [{\citenamefont {Pereira}\ \emph {et~al.}(2009)\citenamefont
  {Pereira}, \citenamefont {Neto},\ and\ \citenamefont {Peres}}]{Pereira2009}%
  \BibitemOpen
  \bibfield  {author} {\bibinfo {author} {\bibfnamefont {V.~M.}\ \bibnamefont
  {Pereira}}, \bibinfo {author} {\bibfnamefont {A.~H.~C.}\ \bibnamefont
  {Neto}},\ and\ \bibinfo {author} {\bibfnamefont {N.~M.~R.}\ \bibnamefont
  {Peres}},\ }\bibfield  {title} {\bibinfo {title} {{Tight-binding approach to
  uniaxial strain in graphene}},\ }\href
  {https://doi.org/10.1103/physrevb.80.045401} {\bibfield  {journal} {\bibinfo
  {journal} {Phys. Rev. B}\ }\textbf {\bibinfo {volume} {80}},\ \bibinfo
  {pages} {045401} (\bibinfo {year} {2009})}\BibitemShut {NoStop}%
\bibitem [{\citenamefont {Naumis}\ \emph {et~al.}(2017)\citenamefont {Naumis},
  \citenamefont {Barraza-Lopez}, \citenamefont {Oliva-Leyva},\ and\
  \citenamefont {Terrones}}]{Naumis2017}%
  \BibitemOpen
  \bibfield  {author} {\bibinfo {author} {\bibfnamefont {G.~G.}\ \bibnamefont
  {Naumis}}, \bibinfo {author} {\bibfnamefont {S.}~\bibnamefont
  {Barraza-Lopez}}, \bibinfo {author} {\bibfnamefont {M.}~\bibnamefont
  {Oliva-Leyva}},\ and\ \bibinfo {author} {\bibfnamefont {H.}~\bibnamefont
  {Terrones}},\ }\bibfield  {title} {\bibinfo {title} {{Electronic and optical
  properties of strained graphene and other strained 2D materials: a review}},\
  }\href {https://doi.org/10.1088/1361-6633/aa74ef} {\bibfield  {journal}
  {\bibinfo  {journal} {Reports on Progress in Physics}\ }\textbf {\bibinfo
  {volume} {80}},\ \bibinfo {pages} {096501} (\bibinfo {year}
  {2017})}\BibitemShut {NoStop}%
\bibitem [{\citenamefont {Mohiuddin}\ \emph {et~al.}(2009)\citenamefont
  {Mohiuddin}, \citenamefont {Lombardo}, \citenamefont {Nair}, \citenamefont
  {Bonetti}, \citenamefont {Savini}, \citenamefont {Jalil}, \citenamefont
  {Bonini}, \citenamefont {Basko}, \citenamefont {Galiotis}, \citenamefont
  {Marzari}, \citenamefont {Novoselov}, \citenamefont {Geim},\ and\
  \citenamefont {Ferrari}}]{Mohiuddin2009}%
  \BibitemOpen
  \bibfield  {author} {\bibinfo {author} {\bibfnamefont {T.~M.~G.}\
  \bibnamefont {Mohiuddin}}, \bibinfo {author} {\bibfnamefont {A.}~\bibnamefont
  {Lombardo}}, \bibinfo {author} {\bibfnamefont {R.~R.}\ \bibnamefont {Nair}},
  \bibinfo {author} {\bibfnamefont {A.}~\bibnamefont {Bonetti}}, \bibinfo
  {author} {\bibfnamefont {G.}~\bibnamefont {Savini}}, \bibinfo {author}
  {\bibfnamefont {R.}~\bibnamefont {Jalil}}, \bibinfo {author} {\bibfnamefont
  {N.}~\bibnamefont {Bonini}}, \bibinfo {author} {\bibfnamefont {D.~M.}\
  \bibnamefont {Basko}}, \bibinfo {author} {\bibfnamefont {C.}~\bibnamefont
  {Galiotis}}, \bibinfo {author} {\bibfnamefont {N.}~\bibnamefont {Marzari}},
  \bibinfo {author} {\bibfnamefont {K.~S.}\ \bibnamefont {Novoselov}}, \bibinfo
  {author} {\bibfnamefont {A.~K.}\ \bibnamefont {Geim}},\ and\ \bibinfo
  {author} {\bibfnamefont {A.~C.}\ \bibnamefont {Ferrari}},\ }\bibfield
  {title} {\bibinfo {title} {{Uniaxial strain in graphene by Raman
  spectroscopy: $G$ peak splitting, Gr\"uneisen parameters, and sample
  orientation}},\ }\href {https://doi.org/10.1103/PhysRevB.79.205433}
  {\bibfield  {journal} {\bibinfo  {journal} {Phys. Rev. B}\ }\textbf {\bibinfo
  {volume} {79}},\ \bibinfo {pages} {205433} (\bibinfo {year}
  {2009})}\BibitemShut {NoStop}%
\bibitem [{\citenamefont {Rold{\'{a}}n}\ \emph {et~al.}(2015)\citenamefont
  {Rold{\'{a}}n}, \citenamefont {Castellanos-Gomez}, \citenamefont
  {Cappelluti},\ and\ \citenamefont {Guinea}}]{Roldan2015}%
  \BibitemOpen
  \bibfield  {author} {\bibinfo {author} {\bibfnamefont {R.}~\bibnamefont
  {Rold{\'{a}}n}}, \bibinfo {author} {\bibfnamefont {A.}~\bibnamefont
  {Castellanos-Gomez}}, \bibinfo {author} {\bibfnamefont {E.}~\bibnamefont
  {Cappelluti}},\ and\ \bibinfo {author} {\bibfnamefont {F.}~\bibnamefont
  {Guinea}},\ }\bibfield  {title} {\bibinfo {title} {Strain engineering in
  semiconducting two-dimensional crystals},\ }\href
  {https://doi.org/10.1088/0953-8984/27/31/313201} {\bibfield  {journal}
  {\bibinfo  {journal} {J. Phys.: Condens. Matter}\ }\textbf {\bibinfo {volume}
  {27}},\ \bibinfo {pages} {313201} (\bibinfo {year} {2015})}\BibitemShut
  {NoStop}%
\bibitem [{\citenamefont {Mili\ifmmode \acute{c}\else
  \'{c}\fi{}evi\ifmmode~\acute{c}\else \'{c}\fi{}}\ \emph
  {et~al.}(2019)\citenamefont {Mili\ifmmode \acute{c}\else
  \'{c}\fi{}evi\ifmmode~\acute{c}\else \'{c}\fi{}}, \citenamefont {Montambaux},
  \citenamefont {Ozawa}, \citenamefont {Jamadi}, \citenamefont {Real},
  \citenamefont {Sagnes}, \citenamefont {Lema\^{\i}tre}, \citenamefont
  {Le~Gratiet}, \citenamefont {Harouri}, \citenamefont {Bloch},\ and\
  \citenamefont {Amo}}]{Montambaux2019}%
  \BibitemOpen
  \bibfield  {author} {\bibinfo {author} {\bibfnamefont {M.}~\bibnamefont
  {Mili\ifmmode \acute{c}\else \'{c}\fi{}evi\ifmmode~\acute{c}\else
  \'{c}\fi{}}}, \bibinfo {author} {\bibfnamefont {G.}~\bibnamefont
  {Montambaux}}, \bibinfo {author} {\bibfnamefont {T.}~\bibnamefont {Ozawa}},
  \bibinfo {author} {\bibfnamefont {O.}~\bibnamefont {Jamadi}}, \bibinfo
  {author} {\bibfnamefont {B.}~\bibnamefont {Real}}, \bibinfo {author}
  {\bibfnamefont {I.}~\bibnamefont {Sagnes}}, \bibinfo {author} {\bibfnamefont
  {A.}~\bibnamefont {Lema\^{\i}tre}}, \bibinfo {author} {\bibfnamefont
  {L.}~\bibnamefont {Le~Gratiet}}, \bibinfo {author} {\bibfnamefont
  {A.}~\bibnamefont {Harouri}}, \bibinfo {author} {\bibfnamefont
  {J.}~\bibnamefont {Bloch}},\ and\ \bibinfo {author} {\bibfnamefont
  {A.}~\bibnamefont {Amo}},\ }\bibfield  {title} {\bibinfo {title} {{Type-III
  and Tilted Dirac Cones Emerging from Flat Bands in Photonic Orbital
  Graphene}},\ }\href {https://doi.org/10.1103/PhysRevX.9.031010} {\bibfield
  {journal} {\bibinfo  {journal} {Phys. Rev. X}\ }\textbf {\bibinfo {volume}
  {9}},\ \bibinfo {pages} {031010} (\bibinfo {year} {2019})}\BibitemShut
  {NoStop}%
\bibitem [{\citenamefont {Goerbig}\ \emph {et~al.}(2008)\citenamefont
  {Goerbig}, \citenamefont {Fuchs}, \citenamefont {Montambaux},\ and\
  \citenamefont {Pi\'echon}}]{Goerbig2008}%
  \BibitemOpen
  \bibfield  {author} {\bibinfo {author} {\bibfnamefont {M.~O.}\ \bibnamefont
  {Goerbig}}, \bibinfo {author} {\bibfnamefont {J.-N.}\ \bibnamefont {Fuchs}},
  \bibinfo {author} {\bibfnamefont {G.}~\bibnamefont {Montambaux}},\ and\
  \bibinfo {author} {\bibfnamefont {F.}~\bibnamefont {Pi\'echon}},\ }\bibfield
  {title} {\bibinfo {title} {{Tilted anisotropic Dirac cones in quinoid-type
  graphene and
  $\ensuremath{\alpha}\text{\ensuremath{-}}{(\text{BEDT-TTF})}_{2}{\text{I}}_{3}$}},\
  }\href {https://doi.org/10.1103/PhysRevB.78.045415} {\bibfield  {journal}
  {\bibinfo  {journal} {Phys. Rev. B}\ }\textbf {\bibinfo {volume} {78}},\
  \bibinfo {pages} {045415} (\bibinfo {year} {2008})}\BibitemShut {NoStop}%
\bibitem [{\citenamefont {Zabolotskiy}\ and\ \citenamefont
  {Lozovik}(2016)}]{Zabolotskiy2016}%
  \BibitemOpen
  \bibfield  {author} {\bibinfo {author} {\bibfnamefont {A.~D.}\ \bibnamefont
  {Zabolotskiy}}\ and\ \bibinfo {author} {\bibfnamefont {Y.~E.}\ \bibnamefont
  {Lozovik}},\ }\bibfield  {title} {\bibinfo {title} {{Strain-induced
  pseudomagnetic field in the Dirac semimetal borophene}},\ }\href
  {https://doi.org/10.1103/PhysRevB.94.165403} {\bibfield  {journal} {\bibinfo
  {journal} {Phys. Rev. B}\ }\textbf {\bibinfo {volume} {94}},\ \bibinfo
  {pages} {165403} (\bibinfo {year} {2016})}\BibitemShut {NoStop}%
\bibitem [{\citenamefont {Sadhukhan}\ and\ \citenamefont
  {Agarwal}(2017)}]{Sadhukhan2017}%
  \BibitemOpen
  \bibfield  {author} {\bibinfo {author} {\bibfnamefont {K.}~\bibnamefont
  {Sadhukhan}}\ and\ \bibinfo {author} {\bibfnamefont {A.}~\bibnamefont
  {Agarwal}},\ }\bibfield  {title} {\bibinfo {title} {{Anisotropic plasmons,
  Friedel oscillations, and screening in $8\text{\ensuremath{-}}Pmmn$
  borophene}},\ }\href {https://doi.org/10.1103/PhysRevB.96.035410} {\bibfield
  {journal} {\bibinfo  {journal} {Phys. Rev. B}\ }\textbf {\bibinfo {volume}
  {96}},\ \bibinfo {pages} {035410} (\bibinfo {year} {2017})}\BibitemShut
  {NoStop}%
\bibitem [{\citenamefont {Sharma}\ \emph {et~al.}(2013)\citenamefont {Sharma},
  \citenamefont {Kotov},\ and\ \citenamefont {Castro~Neto}}]{Sharma2013}%
  \BibitemOpen
  \bibfield  {author} {\bibinfo {author} {\bibfnamefont {A.}~\bibnamefont
  {Sharma}}, \bibinfo {author} {\bibfnamefont {V.~N.}\ \bibnamefont {Kotov}},\
  and\ \bibinfo {author} {\bibfnamefont {A.~H.}\ \bibnamefont {Castro~Neto}},\
  }\bibfield  {title} {\bibinfo {title} {Effect of uniaxial strain on
  ferromagnetic instability and formation of localized magnetic states on
  adatoms in graphene},\ }\href {https://doi.org/10.1103/PhysRevB.87.155431}
  {\bibfield  {journal} {\bibinfo  {journal} {Phys. Rev. B}\ }\textbf {\bibinfo
  {volume} {87}},\ \bibinfo {pages} {155431} (\bibinfo {year}
  {2013})}\BibitemShut {NoStop}%
\bibitem [{\citenamefont {{Del Maestro}}\ \emph {et~al.}(2021)\citenamefont
  {{Del Maestro}}, \citenamefont {Wexler}, \citenamefont {Vanegas},
  \citenamefont {Lakoba},\ and\ \citenamefont {Kotov}}]{DelMaestro2021}%
  \BibitemOpen
  \bibfield  {author} {\bibinfo {author} {\bibfnamefont {A.}~\bibnamefont {{Del
  Maestro}}}, \bibinfo {author} {\bibfnamefont {C.}~\bibnamefont {Wexler}},
  \bibinfo {author} {\bibfnamefont {J.~M.}\ \bibnamefont {Vanegas}}, \bibinfo
  {author} {\bibfnamefont {T.}~\bibnamefont {Lakoba}},\ and\ \bibinfo {author}
  {\bibfnamefont {V.~N.}\ \bibnamefont {Kotov}},\ }\bibfield  {title} {\bibinfo
  {title} {{A} perspective on {C}ollective {P}roperties of {A}toms on 2{D}
  materials},\ }\href {https://doi.org/10.1002/aelm.202100607} {\bibfield
  {journal} {\bibinfo  {journal} {Advanced Electronic Materials}\ }\textbf
  {\bibinfo {volume} {8}},\ \bibinfo {pages} {2100607} (\bibinfo {year}
  {2021})}\BibitemShut {NoStop}%
\bibitem [{\citenamefont {Sengupta}\ \emph {et~al.}(2018)\citenamefont
  {Sengupta}, \citenamefont {Nichols}, \citenamefont {{Del Maestro}},\ and\
  \citenamefont {Kotov}}]{Sengupta2018}%
  \BibitemOpen
  \bibfield  {author} {\bibinfo {author} {\bibfnamefont {S.}~\bibnamefont
  {Sengupta}}, \bibinfo {author} {\bibfnamefont {N.~S.}\ \bibnamefont
  {Nichols}}, \bibinfo {author} {\bibfnamefont {A.}~\bibnamefont {{Del
  Maestro}}},\ and\ \bibinfo {author} {\bibfnamefont {V.~N.}\ \bibnamefont
  {Kotov}},\ }\bibfield  {title} {\bibinfo {title} {Theory of liquid film
  growth and wetting instabilities on graphene},\ }\href
  {https://doi.org/10.1103/PhysRevLett.120.236802} {\bibfield  {journal}
  {\bibinfo  {journal} {Phys. Rev. Lett.}\ }\textbf {\bibinfo {volume} {120}},\
  \bibinfo {pages} {236802} (\bibinfo {year} {2018})}\BibitemShut {NoStop}%
\bibitem [{\citenamefont {Kim}\ \emph {et~al.}(2022)\citenamefont {Kim},
  \citenamefont {Elsayed}, \citenamefont {Nichols}, \citenamefont {Lakoba},
  \citenamefont {Vanegas}, \citenamefont {Wexler}, \citenamefont {Kotov},\ and\
  \citenamefont {Maestro}}]{Kim22}%
  \BibitemOpen
  \bibfield  {author} {\bibinfo {author} {\bibfnamefont {S.~W.}\ \bibnamefont
  {Kim}}, \bibinfo {author} {\bibfnamefont {M.}~\bibnamefont {Elsayed}},
  \bibinfo {author} {\bibfnamefont {N.~S.}\ \bibnamefont {Nichols}}, \bibinfo
  {author} {\bibfnamefont {T.}~\bibnamefont {Lakoba}}, \bibinfo {author}
  {\bibfnamefont {J.}~\bibnamefont {Vanegas}}, \bibinfo {author} {\bibfnamefont
  {C.}~\bibnamefont {Wexler}}, \bibinfo {author} {\bibfnamefont {V.~N.}\
  \bibnamefont {Kotov}},\ and\ \bibinfo {author} {\bibfnamefont {A.~D.}\
  \bibnamefont {Maestro}},\ }\href@noop {} {\bibinfo {title} {Strain-induced
  superfluid transition for atoms on graphene}} (\bibinfo {year} {2022}),\
  \Eprint {https://arxiv.org/abs/arXiv:2211.07672} {arXiv:2211.07672}
  \BibitemShut {NoStop}%
\bibitem [{\citenamefont {Simion}\ and\ \citenamefont
  {Giuliani}(2005)}]{Simion2005}%
  \BibitemOpen
  \bibfield  {author} {\bibinfo {author} {\bibfnamefont {G.~E.}\ \bibnamefont
  {Simion}}\ and\ \bibinfo {author} {\bibfnamefont {G.~F.}\ \bibnamefont
  {Giuliani}},\ }\bibfield  {title} {\bibinfo {title} {{Friedel oscillations in
  a Fermi liquid}},\ }\href {https://doi.org/10.1103/PhysRevB.72.045127}
  {\bibfield  {journal} {\bibinfo  {journal} {Phys. Rev. B}\ }\textbf {\bibinfo
  {volume} {72}},\ \bibinfo {pages} {045127} (\bibinfo {year}
  {2005})}\BibitemShut {NoStop}%
\bibitem [{\citenamefont {Giuliani}\ and\ \citenamefont
  {Vignale}(2005)}]{Giuliani2005}%
  \BibitemOpen
  \bibfield  {author} {\bibinfo {author} {\bibfnamefont {G.}~\bibnamefont
  {Giuliani}}\ and\ \bibinfo {author} {\bibfnamefont {G.}~\bibnamefont
  {Vignale}},\ }\href {https://doi.org/10.1017/CBO9780511619915} {\emph
  {\bibinfo {title} {Quantum Theory of the Electron Liquid}}}\ (\bibinfo
  {publisher} {Cambridge University Press},\ \bibinfo {year}
  {2005})\BibitemShut {NoStop}%
\bibitem [{\citenamefont {Cheianov}\ and\ \citenamefont
  {Fal'ko}(2006)}]{Cheianov2006}%
  \BibitemOpen
  \bibfield  {author} {\bibinfo {author} {\bibfnamefont {V.~V.}\ \bibnamefont
  {Cheianov}}\ and\ \bibinfo {author} {\bibfnamefont {V.~I.}\ \bibnamefont
  {Fal'ko}},\ }\bibfield  {title} {\bibinfo {title} {Friedel oscillations,
  impurity scattering, and temperature dependence of resistivity in graphene},\
  }\href {https://doi.org/10.1103/PhysRevLett.97.226801} {\bibfield  {journal}
  {\bibinfo  {journal} {Phys. Rev. Lett.}\ }\textbf {\bibinfo {volume} {97}},\
  \bibinfo {pages} {226801} (\bibinfo {year} {2006})}\BibitemShut {NoStop}%
\bibitem [{\citenamefont {Wunsch}\ \emph {et~al.}(2006)\citenamefont {Wunsch},
  \citenamefont {Stauber}, \citenamefont {Sols},\ and\ \citenamefont
  {Guinea}}]{Wunsch2006}%
  \BibitemOpen
  \bibfield  {author} {\bibinfo {author} {\bibfnamefont {B.}~\bibnamefont
  {Wunsch}}, \bibinfo {author} {\bibfnamefont {T.}~\bibnamefont {Stauber}},
  \bibinfo {author} {\bibfnamefont {F.}~\bibnamefont {Sols}},\ and\ \bibinfo
  {author} {\bibfnamefont {F.}~\bibnamefont {Guinea}},\ }\bibfield  {title}
  {\bibinfo {title} {Dynamical polarization of graphene at finite doping},\
  }\href {https://doi.org/10.1088/1367-2630/8/12/318} {\bibfield  {journal}
  {\bibinfo  {journal} {New Journal of Physics}\ }\textbf {\bibinfo {volume}
  {8}},\ \bibinfo {pages} {318} (\bibinfo {year} {2006})}\BibitemShut {NoStop}%
\bibitem [{\citenamefont {Milstein}\ and\ \citenamefont
  {Terekhov}(2010)}]{Milstein2010}%
  \BibitemOpen
  \bibfield  {author} {\bibinfo {author} {\bibfnamefont {A.~I.}\ \bibnamefont
  {Milstein}}\ and\ \bibinfo {author} {\bibfnamefont {I.~S.}\ \bibnamefont
  {Terekhov}},\ }\bibfield  {title} {\bibinfo {title} {Induced charge generated
  by a potential well in graphene},\ }\href
  {https://doi.org/10.1103/PhysRevB.81.125419} {\bibfield  {journal} {\bibinfo
  {journal} {Phys. Rev. B}\ }\textbf {\bibinfo {volume} {81}},\ \bibinfo
  {pages} {125419} (\bibinfo {year} {2010})}\BibitemShut {NoStop}%
\bibitem [{\citenamefont {Mkhitaryan}\ and\ \citenamefont
  {Mishchenko}(2012)}]{Mkhitaryan2012}%
  \BibitemOpen
  \bibfield  {author} {\bibinfo {author} {\bibfnamefont {V.~V.}\ \bibnamefont
  {Mkhitaryan}}\ and\ \bibinfo {author} {\bibfnamefont {E.~G.}\ \bibnamefont
  {Mishchenko}},\ }\bibfield  {title} {\bibinfo {title} {{Resonant finite-size
  impurities in graphene, unitary limit, and Friedel oscillations}},\ }\href
  {https://doi.org/10.1103/PhysRevB.86.115442} {\bibfield  {journal} {\bibinfo
  {journal} {Phys. Rev. B}\ }\textbf {\bibinfo {volume} {86}},\ \bibinfo
  {pages} {115442} (\bibinfo {year} {2012})}\BibitemShut {NoStop}%
\bibitem [{\citenamefont {VandeVondele}\ \emph {et~al.}(2005)\citenamefont
  {VandeVondele}, \citenamefont {Krack}, \citenamefont {Mohamed}, \citenamefont
  {Parrinello}, \citenamefont {Chassaing},\ and\ \citenamefont
  {Hutter}}]{Quickstep}%
  \BibitemOpen
  \bibfield  {author} {\bibinfo {author} {\bibfnamefont {J.}~\bibnamefont
  {VandeVondele}}, \bibinfo {author} {\bibfnamefont {M.}~\bibnamefont {Krack}},
  \bibinfo {author} {\bibfnamefont {F.}~\bibnamefont {Mohamed}}, \bibinfo
  {author} {\bibfnamefont {M.}~\bibnamefont {Parrinello}}, \bibinfo {author}
  {\bibfnamefont {T.}~\bibnamefont {Chassaing}},\ and\ \bibinfo {author}
  {\bibfnamefont {J.}~\bibnamefont {Hutter}},\ }\bibfield  {title} {\bibinfo
  {title} {{Quickstep: Fast and accurate density functional calculations using
  a mixed Gaussian and plane waves approach}},\ }\href
  {https://doi.org/https://doi.org/10.1016/j.cpc.2004.12.014} {\bibfield
  {journal} {\bibinfo  {journal} {Comput. Phys. Commun.}\ }\textbf {\bibinfo
  {volume} {167}},\ \bibinfo {pages} {103} (\bibinfo {year}
  {2005})}\BibitemShut {NoStop}%
\bibitem [{\citenamefont {Kühne}\ \emph {et~al.}(2020)\citenamefont {Kühne},
  \citenamefont {Iannuzzi}, \citenamefont {Del~Ben}, \citenamefont {Rybkin},
  \citenamefont {Seewald}, \citenamefont {Stein}, \citenamefont {Laino},
  \citenamefont {Khaliullin}, \citenamefont {Schütt}, \citenamefont
  {Schiffmann}, \citenamefont {Golze}, \citenamefont {Wilhelm}, \citenamefont
  {Chulkov}, \citenamefont {Bani-Hashemian}, \citenamefont {Weber},
  \citenamefont {Borštnik}, \citenamefont {Taillefumier}, \citenamefont
  {Jakobovits}, \citenamefont {Lazzaro}, \citenamefont {Pabst}, \citenamefont
  {Müller}, \citenamefont {Schade}, \citenamefont {Guidon}, \citenamefont
  {Andermatt}, \citenamefont {Holmberg}, \citenamefont {Schenter},
  \citenamefont {Hehn}, \citenamefont {Bussy}, \citenamefont {Belleflamme},
  \citenamefont {Tabacchi}, \citenamefont {Glöß}, \citenamefont {Lass},
  \citenamefont {Bethune}, \citenamefont {Mundy}, \citenamefont {Plessl},
  \citenamefont {Watkins}, \citenamefont {VandeVondele}, \citenamefont
  {Krack},\ and\ \citenamefont {Hutter}}]{CP2Krev}%
  \BibitemOpen
  \bibfield  {author} {\bibinfo {author} {\bibfnamefont {T.~D.}\ \bibnamefont
  {Kühne}}, \bibinfo {author} {\bibfnamefont {M.}~\bibnamefont {Iannuzzi}},
  \bibinfo {author} {\bibfnamefont {M.}~\bibnamefont {Del~Ben}}, \bibinfo
  {author} {\bibfnamefont {V.~V.}\ \bibnamefont {Rybkin}}, \bibinfo {author}
  {\bibfnamefont {P.}~\bibnamefont {Seewald}}, \bibinfo {author} {\bibfnamefont
  {F.}~\bibnamefont {Stein}}, \bibinfo {author} {\bibfnamefont
  {T.}~\bibnamefont {Laino}}, \bibinfo {author} {\bibfnamefont {R.~Z.}\
  \bibnamefont {Khaliullin}}, \bibinfo {author} {\bibfnamefont
  {O.}~\bibnamefont {Schütt}}, \bibinfo {author} {\bibfnamefont
  {F.}~\bibnamefont {Schiffmann}}, \bibinfo {author} {\bibfnamefont
  {D.}~\bibnamefont {Golze}}, \bibinfo {author} {\bibfnamefont
  {J.}~\bibnamefont {Wilhelm}}, \bibinfo {author} {\bibfnamefont
  {S.}~\bibnamefont {Chulkov}}, \bibinfo {author} {\bibfnamefont {M.~H.}\
  \bibnamefont {Bani-Hashemian}}, \bibinfo {author} {\bibfnamefont
  {V.}~\bibnamefont {Weber}}, \bibinfo {author} {\bibfnamefont
  {U.}~\bibnamefont {Borštnik}}, \bibinfo {author} {\bibfnamefont
  {M.}~\bibnamefont {Taillefumier}}, \bibinfo {author} {\bibfnamefont {A.~S.}\
  \bibnamefont {Jakobovits}}, \bibinfo {author} {\bibfnamefont
  {A.}~\bibnamefont {Lazzaro}}, \bibinfo {author} {\bibfnamefont
  {H.}~\bibnamefont {Pabst}}, \bibinfo {author} {\bibfnamefont
  {T.}~\bibnamefont {Müller}}, \bibinfo {author} {\bibfnamefont
  {R.}~\bibnamefont {Schade}}, \bibinfo {author} {\bibfnamefont
  {M.}~\bibnamefont {Guidon}}, \bibinfo {author} {\bibfnamefont
  {S.}~\bibnamefont {Andermatt}}, \bibinfo {author} {\bibfnamefont
  {N.}~\bibnamefont {Holmberg}}, \bibinfo {author} {\bibfnamefont {G.~K.}\
  \bibnamefont {Schenter}}, \bibinfo {author} {\bibfnamefont {A.}~\bibnamefont
  {Hehn}}, \bibinfo {author} {\bibfnamefont {A.}~\bibnamefont {Bussy}},
  \bibinfo {author} {\bibfnamefont {F.}~\bibnamefont {Belleflamme}}, \bibinfo
  {author} {\bibfnamefont {G.}~\bibnamefont {Tabacchi}}, \bibinfo {author}
  {\bibfnamefont {A.}~\bibnamefont {Glöß}}, \bibinfo {author} {\bibfnamefont
  {M.}~\bibnamefont {Lass}}, \bibinfo {author} {\bibfnamefont {I.}~\bibnamefont
  {Bethune}}, \bibinfo {author} {\bibfnamefont {C.~J.}\ \bibnamefont {Mundy}},
  \bibinfo {author} {\bibfnamefont {C.}~\bibnamefont {Plessl}}, \bibinfo
  {author} {\bibfnamefont {M.}~\bibnamefont {Watkins}}, \bibinfo {author}
  {\bibfnamefont {J.}~\bibnamefont {VandeVondele}}, \bibinfo {author}
  {\bibfnamefont {M.}~\bibnamefont {Krack}},\ and\ \bibinfo {author}
  {\bibfnamefont {J.}~\bibnamefont {Hutter}},\ }\bibfield  {title} {\bibinfo
  {title} {{CP2K: An electronic structure and molecular dynamics software
  package - Quickstep: Efficient and accurate electronic structure
  calculations}},\ }\href {https://doi.org/10.1063/5.0007045} {\bibfield
  {journal} {\bibinfo  {journal} {J. Chem. Phys.}\ }\textbf {\bibinfo {volume}
  {152}},\ \bibinfo {pages} {194103} (\bibinfo {year} {2020})}\BibitemShut
  {NoStop}%
\bibitem [{\citenamefont {Perdew}\ \emph {et~al.}(2008)\citenamefont {Perdew},
  \citenamefont {Ruzsinszky}, \citenamefont {Csonka}, \citenamefont {Vydrov},
  \citenamefont {Scuseria}, \citenamefont {Constantin}, \citenamefont {Zhou},\
  and\ \citenamefont {Burke}}]{PBESol}%
  \BibitemOpen
  \bibfield  {author} {\bibinfo {author} {\bibfnamefont {J.~P.}\ \bibnamefont
  {Perdew}}, \bibinfo {author} {\bibfnamefont {A.}~\bibnamefont {Ruzsinszky}},
  \bibinfo {author} {\bibfnamefont {G.~I.}\ \bibnamefont {Csonka}}, \bibinfo
  {author} {\bibfnamefont {O.~A.}\ \bibnamefont {Vydrov}}, \bibinfo {author}
  {\bibfnamefont {G.~E.}\ \bibnamefont {Scuseria}}, \bibinfo {author}
  {\bibfnamefont {L.~A.}\ \bibnamefont {Constantin}}, \bibinfo {author}
  {\bibfnamefont {X.}~\bibnamefont {Zhou}},\ and\ \bibinfo {author}
  {\bibfnamefont {K.}~\bibnamefont {Burke}},\ }\bibfield  {title} {\bibinfo
  {title} {Restoring the density-gradient expansion for exchange in solids and
  surfaces},\ }\href {https://link.aps.org/doi/10.1103/PhysRevLett.100.136406}
  {\bibfield  {journal} {\bibinfo  {journal} {Phys. Rev. Lett.}\ }\textbf
  {\bibinfo {volume} {100}},\ \bibinfo {pages} {136406} (\bibinfo {year}
  {2008})}\BibitemShut {NoStop}%
\bibitem [{\citenamefont {Perdew}\ \emph {et~al.}(1996)\citenamefont {Perdew},
  \citenamefont {Burke},\ and\ \citenamefont {Ernzerhof}}]{PBE}%
  \BibitemOpen
  \bibfield  {author} {\bibinfo {author} {\bibfnamefont {J.~P.}\ \bibnamefont
  {Perdew}}, \bibinfo {author} {\bibfnamefont {K.}~\bibnamefont {Burke}},\ and\
  \bibinfo {author} {\bibfnamefont {M.}~\bibnamefont {Ernzerhof}},\ }\bibfield
  {title} {\bibinfo {title} {Generalized gradient approximation made simple},\
  }\href {https://doi.org/10.1103/PhysRevLett.77.3865} {\bibfield  {journal}
  {\bibinfo  {journal} {Phys. Rev. Lett.}\ }\textbf {\bibinfo {volume} {77}},\
  \bibinfo {pages} {3865} (\bibinfo {year} {1996})}\BibitemShut {NoStop}%
\bibitem [{\citenamefont {VandeVondele}\ and\ \citenamefont
  {Hutter}(2003)}]{CP2K_OT}%
  \BibitemOpen
  \bibfield  {author} {\bibinfo {author} {\bibfnamefont {J.}~\bibnamefont
  {VandeVondele}}\ and\ \bibinfo {author} {\bibfnamefont {J.}~\bibnamefont
  {Hutter}},\ }\bibfield  {title} {\bibinfo {title} {An efficient orbital
  transformation method for electronic structure calculations},\ }\href
  {https://doi.org/10.1063/1.1543154} {\bibfield  {journal} {\bibinfo
  {journal} {J. Chem. Phys.}\ }\textbf {\bibinfo {volume} {118}},\ \bibinfo
  {pages} {4365} (\bibinfo {year} {2003})}\BibitemShut {NoStop}%
\bibitem [{\citenamefont {VandeVondele}\ and\ \citenamefont
  {Hutter}(2007)}]{MOLOPT}%
  \BibitemOpen
  \bibfield  {author} {\bibinfo {author} {\bibfnamefont {J.}~\bibnamefont
  {VandeVondele}}\ and\ \bibinfo {author} {\bibfnamefont {J.}~\bibnamefont
  {Hutter}},\ }\bibfield  {title} {\bibinfo {title} {{Gaussian basis sets for
  accurate calculations on molecular systems in gas and condensed phases}},\
  }\href {https://doi.org/10.1063/1.2770708} {\bibfield  {journal} {\bibinfo
  {journal} {The Journal of Chemical Physics}\ }\textbf {\bibinfo {volume}
  {127}},\ \bibinfo {pages} {114105} (\bibinfo {year} {2007})}\BibitemShut
  {NoStop}%
\bibitem [{\citenamefont {Goedecker}\ \emph {et~al.}(1996)\citenamefont
  {Goedecker}, \citenamefont {Teter},\ and\ \citenamefont {Hutter}}]{GTH}%
  \BibitemOpen
  \bibfield  {author} {\bibinfo {author} {\bibfnamefont {S.}~\bibnamefont
  {Goedecker}}, \bibinfo {author} {\bibfnamefont {M.}~\bibnamefont {Teter}},\
  and\ \bibinfo {author} {\bibfnamefont {J.}~\bibnamefont {Hutter}},\
  }\bibfield  {title} {\bibinfo {title} {{Separable dual-space Gaussian
  pseudopotentials}},\ }\href {https://doi.org/10.1103/PhysRevB.54.1703}
  {\bibfield  {journal} {\bibinfo  {journal} {Phys. Rev. B}\ }\textbf {\bibinfo
  {volume} {54}},\ \bibinfo {pages} {1703} (\bibinfo {year}
  {1996})}\BibitemShut {NoStop}%
\bibitem [{\citenamefont {Hartwigsen}\ \emph {et~al.}(1998)\citenamefont
  {Hartwigsen}, \citenamefont {Goedecker},\ and\ \citenamefont
  {Hutter}}]{GTH2}%
  \BibitemOpen
  \bibfield  {author} {\bibinfo {author} {\bibfnamefont {C.}~\bibnamefont
  {Hartwigsen}}, \bibinfo {author} {\bibfnamefont {S.}~\bibnamefont
  {Goedecker}},\ and\ \bibinfo {author} {\bibfnamefont {J.}~\bibnamefont
  {Hutter}},\ }\bibfield  {title} {\bibinfo {title} {{Relativistic separable
  dual-space Gaussian pseudopotentials from H to Rn}},\ }\href
  {https://doi.org/10.1103/PhysRevB.58.3641} {\bibfield  {journal} {\bibinfo
  {journal} {Phys. Rev. B}\ }\textbf {\bibinfo {volume} {58}},\ \bibinfo
  {pages} {3641} (\bibinfo {year} {1998})}\BibitemShut {NoStop}%
\bibitem [{\citenamefont {Krack}(2005)}]{GTH3}%
  \BibitemOpen
  \bibfield  {author} {\bibinfo {author} {\bibfnamefont {M.}~\bibnamefont
  {Krack}},\ }\bibfield  {title} {\bibinfo {title} {{Pseudopotentials for H to
  Kr optimized for gradient-corrected exchange-correlation functionals}},\
  }\href {https://doi.org/10.1007/s00214-005-0655-y} {\bibfield  {journal}
  {\bibinfo  {journal} {Theor Chem Account}\ }\textbf {\bibinfo {volume}
  {114}},\ \bibinfo {pages} {145} (\bibinfo {year} {2005})}\BibitemShut
  {NoStop}%
\bibitem [{\citenamefont {Yekta}\ \emph {et~al.}(2023)\citenamefont {Yekta},
  \citenamefont {Hadipour},\ and\ \citenamefont {Jafari}}]{Yekta2023}%
  \BibitemOpen
  \bibfield  {author} {\bibinfo {author} {\bibfnamefont {Y.}~\bibnamefont
  {Yekta}}, \bibinfo {author} {\bibfnamefont {H.}~\bibnamefont {Hadipour}},\
  and\ \bibinfo {author} {\bibfnamefont {S.~A.}\ \bibnamefont {Jafari}},\
  }\bibfield  {title} {\bibinfo {title} {{Tunning the tilt of the Dirac cone by
  atomic manipulations in 8Pmmn borophene}},\ }\bibfield  {journal} {\bibinfo
  {journal} {Communications Physics}\ }\textbf {\bibinfo {volume} {6}},\ \href
  {https://doi.org/10.1038/s42005-023-01161-9} {10.1038/s42005-023-01161-9}
  (\bibinfo {year} {2023})\BibitemShut {NoStop}%
\bibitem [{\citenamefont {Radovi\ifmmode~\acute{c}\else \'{c}\fi{}}\ \emph
  {et~al.}(2008)\citenamefont {Radovi\ifmmode~\acute{c}\else \'{c}\fi{}},
  \citenamefont {Had\ifmmode~\check{z}\else \v{z}\fi{}ievski},\ and\
  \citenamefont {Mi\ifmmode \check{s}\else
  \v{s}\fi{}kovi\ifmmode~\acute{c}\else \'{c}\fi{}}}]{Radovic2008}%
  \BibitemOpen
  \bibfield  {author} {\bibinfo {author} {\bibfnamefont {I.}~\bibnamefont
  {Radovi\ifmmode~\acute{c}\else \'{c}\fi{}}}, \bibinfo {author} {\bibfnamefont
  {L.}~\bibnamefont {Had\ifmmode~\check{z}\else \v{z}\fi{}ievski}},\ and\
  \bibinfo {author} {\bibfnamefont {Z.~L.}\ \bibnamefont {Mi\ifmmode
  \check{s}\else \v{s}\fi{}kovi\ifmmode~\acute{c}\else \'{c}\fi{}}},\
  }\bibfield  {title} {\bibinfo {title} {Polarization of supported graphene by
  slowly moving charges},\ }\href {https://doi.org/10.1103/PhysRevB.77.075428}
  {\bibfield  {journal} {\bibinfo  {journal} {Phys. Rev. B}\ }\textbf {\bibinfo
  {volume} {77}},\ \bibinfo {pages} {075428} (\bibinfo {year}
  {2008})}\BibitemShut {NoStop}%
\bibitem [{\citenamefont {Pyatkovskiy}(2008)}]{Pyatkovskiy2009}%
  \BibitemOpen
  \bibfield  {author} {\bibinfo {author} {\bibfnamefont {P.~K.}\ \bibnamefont
  {Pyatkovskiy}},\ }\bibfield  {title} {\bibinfo {title} {Dynamical
  polarization, screening, and plasmons in gapped graphene},\ }\href
  {https://doi.org/10.1088/0953-8984/21/2/025506} {\bibfield  {journal}
  {\bibinfo  {journal} {Journal of Physics: Condensed Matter}\ }\textbf
  {\bibinfo {volume} {21}},\ \bibinfo {pages} {025506} (\bibinfo {year}
  {2008})}\BibitemShut {NoStop}%
\end{thebibliography}%

\end{document}